\documentclass[lettersize,journal]{IEEEtran}
\usepackage{amsmath,amsfonts}
\usepackage{algorithm}
\usepackage{array}
\usepackage[caption=false,font=normalsize,labelfont=sf,textfont=sf]{subfig}
\usepackage{textcomp}
\usepackage{stfloats}
\usepackage{url}
\usepackage{verbatim}
\usepackage{graphicx}
\usepackage{cite}
\usepackage{bbding}
\usepackage{multirow}
\usepackage{enumitem}
\usepackage{graphicx}
\usepackage{algpseudocode}
\usepackage{booktabs}
\usepackage{caption}
\usepackage{xspace}
\usepackage[commandnameprefix=always]{changes}
\usepackage[normalem]{ulem}
\useunder{\uline}{\ul}{}

\newcommand{\methodName}{HAGI++\xspace}
\newcommand{\methodNameOld}{HAGI\xspace}

\newcommand\change[1]{\textcolor{black}{#1}}
\newcommand\revise[1]{\textcolor{black}{#1}}
\usepackage{etoolbox}

\begin{document}

\title{\methodName: Head-Assisted Gaze Imputation and Generation}

\author{Chuhan~Jiao,~Zhiming~Hu{}$^{\ast}$,~Andreas~Bulling
\IEEEcompsocitemizethanks{\IEEEcompsocthanksitem 
Chuhan Jiao and Andreas Bulling are with the Institute for Visualisation and Interactive Systems, University of Stuttgart, Germany. \protect E-mail: \{chuhan.jiao, andreas.bulling\}@vis.uni-stuttgart.de.
Zhiming Hu is with The Hong Kong University of Science and Technology (Guangzhou), China, and The Hong Kong University of Science and Technology, China. \protect E-mail: zhiminghu@hkust-gz.edu.cn.

{}$^{\ast}$Corresponding author
}}

\markboth{Journal of \LaTeX\ Class Files,~Vol.~14, No.~8, August~2021}%
{Shell \MakeLowercase{\textit{et al.}}: A Sample Article Using IEEEtran.cls for IEEE Journals}


\maketitle

\begin{figure*}[t]
    \centering
    \vspace{-2em}
    \includegraphics[width=0.8\linewidth]{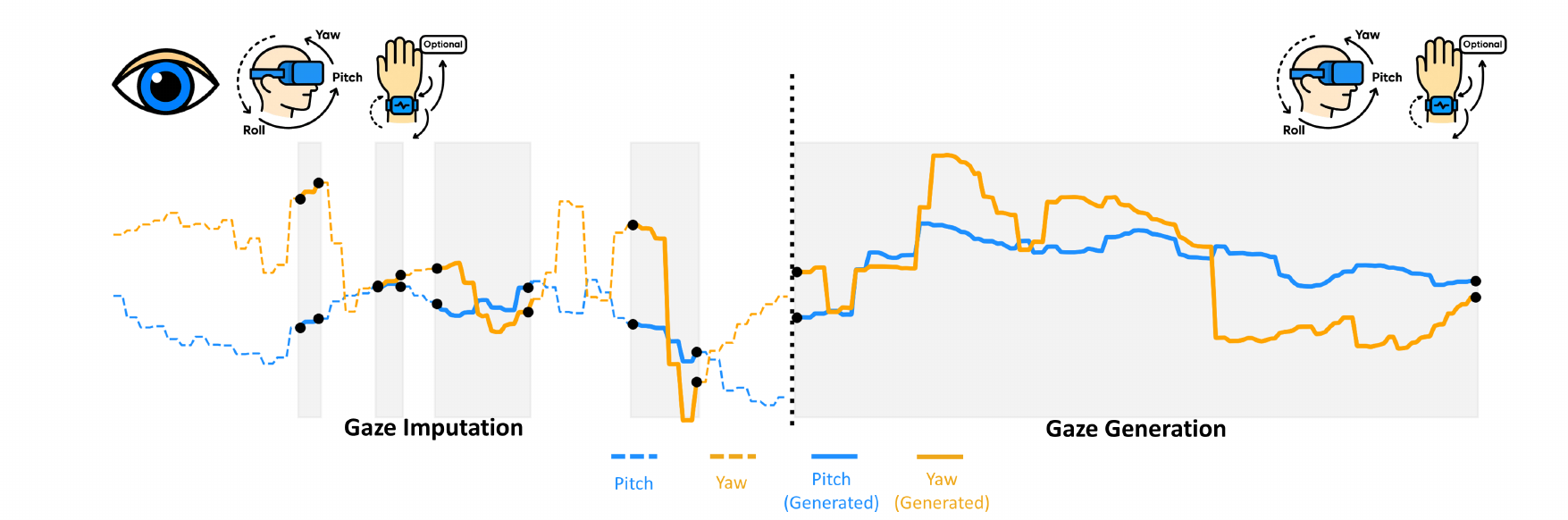}
    \caption{
        Missing data is inevitable in mobile eye tracking. 
        \methodName{} is a novel multi-modal diffusion model for gaze data imputation that leverages the close coordination between eye and head movements. 
        The input to our method is gaze data with missing values and time-aligned head movements captured by sensors readily available in mobile eye trackers (left). 
        For XR headsets without built-in eye tracking, \methodName{} can generate gaze data directly from head movements (right). 
        Furthermore, when wrist or hand motion data from commodity wearable devices are available, \methodName{} can exploit eye-hand-head coordination for both tasks. 
        \methodName{} achieves lower mean angular error and produces more realistic gaze trajectories than previous methods.
    }
    \label{fig:teaser}
    \vspace{-1em}
\end{figure*}

\begin{abstract}
Mobile eye‑tracking is crucial for capturing human visual attention in real‑world and XR settings, supporting research and human‑computer interaction. Yet blinks, pupil‑detection errors and lighting changes create missing values that hinder gaze analysis.
We present \methodName{}, a multi‑modal diffusion‑based imputation method that, for the first time, leverages integrated head‑orientation sensors to exploit the natural correlation between head and eye movements.
\change{Using a transformer‑based diffusion model, it learns cross‑modal dependencies between eye and head data and can additionally incorporate wrist/hand motion when such wearable signals are available.}
Evaluations on the large‑scale Nymeria, Ego‑Exo4D and HOT3D datasets show that \methodName{} consistently outperforms traditional interpolation and deep‑learning time‑series imputation baselines.
Statistical analysis confirms that its gaze‑velocity distributions closely match real human behaviour, yielding realistic imputations.
Even when 100\% of gaze data are missing (pure gaze generation), \methodName{} exceeds methods that rely on the visual inputs and the methods rely on full‑body motion capture by incorporating wrist motion from commercial wearables.
Our approach enables more complete, accurate eye‑gaze recordings in real‑world contexts, enhancing gaze‑based analysis and interaction across many applications.

\end{abstract}


\section{Introduction}

Mobile eye tracking has become an essential tool for studying human behaviour \cite{steil15_ubicomp, hu2022ehtask} and has emerged as an attractive modality for interaction in real-world \cite{lystbaek2024hands, turkmen2024eyeguide} and virtual environments \cite{kaplanyan2019deepfovea,deng2022fov}. 
Also, latest commercial head-mounted extended reality (XR) devices, such as the Apple Vision Pro, Meta Quest Pro, Microsoft HoloLens 2, or the Project Aria Glasses \cite{engel2023project} are equipped with integrated eye tracking functionality.

A fundamental challenge in working with gaze data recorded using mobile eye trackers is the prevalence of missing values arising from blinks, pupil detection failures, occlusions, or illumination changes \cite{nystrom2025fundamentals, nystrom2013influence, blignaut2014eye}. 
Missing values can significantly degrade data quality and, in the worst case, render gaze recordings unusable for further analysis and downstream applications~\cite{grootjen2023highlighting}.  
Prior research has largely adopted two strategies to address this issue: discarding missing values entirely~\cite{jiao23_uist, nakano2010estimating, ishii2013gaze} or imputing them using interpolation methods, such as linear~\cite{bulling2008s, huang2019saccalib} or nearest neighbour interpolation~\cite{zheng2022identification}. 
While discarding data ensures data integrity, it results in discontinuities, making gaze data unsuitable for applications requiring temporal completeness and consistency, such as for training machine learning models~\cite{hu2020dgaze,hu2021fixationnet,hu2022ehtask}. 
Data imputation preserves continuity but fails to accurately reconstruct naturalistic gaze trajectories or match the velocity profile of real human eye movements~\cite{grootjen2023highlighting}. 
As such, none of these existing approaches is fully satisfying
and there remains a critical need for advanced gaze imputation techniques that can handle missing gaze data in a robust, continuity-preserving, and biologically plausible manner.  

We introduce \methodName~-- a novel multi-modal approach that leverages the close coordination between eye and head movements (also known as eye-head coordination) for gaze imputation \footnote{Our code is available at\newline \url{https://git.cai.simtech.uni-stuttgart.de/public-projects/HAGI}}.
Our method exploits the fact that along with gaze data, latest mobile eye trackers and XR headsets are also readily equipped with sensors for head tracking, such as inertial sensors or vision-based approaches for self-localisation and mapping.
Unlike prior methods that treat gaze as a standalone signal, \methodName is the first to integrate head movement information to infer missing gaze values.
In addition, recent mobile eye trackers and XR systems are increasingly paired with wearable devices such as controllers or wristbands, which capture hand or wrist movements for multi-modal interaction based on natural eye-hand coordination. \methodName{} is readily extensible to such inputs, allowing us to exploit eye--head--hand coordination for gaze imputation. 
\change{At the core of \methodName{} is a transformer-based diffusion model that captures cross-modal dependencies between eye and head representations, with an optional extension to wrist/hand signals for modelling gaze--wrist coordination when such data are available.}

We evaluate \methodName on head-assisted gaze imputation on three large-scale gaze datasets that cover different everyday indoor and outdoor activities: Nymeria~\cite{ma2024nymeria}, Ego-Exo4D~\cite{grauman2024ego}, and HOT3D~\cite{banerjee2024hot3d}.
Results of these evaluations show that our approach significantly outperforms both interpolation-based methods and deep learning-based time-series imputation baselines.
Our method reduces mean angular error (MAE) by up to 25.3\%, and yields gaze velocity distributions that more closely resemble real human gaze movements than other methods.
Furthermore, using the wrist-movement signals available in Nymeria, we demonstrate that incorporating wrist movements further improves gaze imputation in high data-loss scenarios and achieves performance on par with head-only input under low data-loss conditions.
Finally, we evaluate \methodName{} under the extreme setting of 100\% gaze data loss (pure gaze generation), including both 3D gaze generation and egocentric gaze estimation from the generated gaze directions.
\methodName{} outperforms EgoM2P \cite{li2025egom2p}, an egocentric foundation model that relies on visual inputs for gaze estimation, achieving lower MAE and MSE using only head-motion input.
Additionally, using only commodity wearable inputs, \methodName{} outperforms the previous state-of-the-art gaze generation method~\cite{hu24pose2gaze}, which requires full-body motion capture, by 17.6\% on the Nymeria dataset.
By explicitly modelling eye-head (and, if available, eye-head-hand) coordination, \methodName enables more reliable gaze-based analysis and interaction using mobile eye tracking.
Our approach enhances the quality of gaze data in real-world scenarios, making it particularly beneficial for applications in XR, behavioural research, and gaze-based human-computer interaction.
Furthermore, by supporting plausible gaze data generation from commodity wearables, \methodName opens new opportunities for computer graphics and interactive systems where realistic gaze behaviour is a critical component.  

An earlier version of this work was published at ACM UIST 2025~\cite{jiao25_uist}. This paper substantially extends that version in several key aspects:
\begin{enumerate}[leftmargin = *]
    \item \textbf{New model architecture.} We introduce a novel architecture, \methodName, that consistently outperforms our previous method and offers faster inference.
    
    \item \textbf{Optional integration of wrist movement signals.} We extend the framework to incorporate wrist signals from commodity wearables, showing that these cues improve robustness under severe data loss and highlight the value of multi-modal eye–hand–head coordination. 
    
    \item \textbf{Evaluation under 100\% gaze data loss.} We present new experiments in the extreme case of 100\% missing gaze data, including 3D gaze generation and egocentric gaze estimation from the generated gaze directions. 
    
    \item \textbf{Comprehensive ablation analysis.} We perform detailed ablation studies to quantify the individual contributions of head rotation, head translation, wrist rotation, and wrist translation to gaze imputation performance.  
\end{enumerate}

\section{Related Work}

\subsection{Gaze Imputation}

The task of gaze imputation, which involves reconstructing missing values within recorded gaze data, remains an underexplored area of research. The predominant approach to handle missing values in most prior studies has been to exclude instances containing missing gaze data entirely \cite{jiao23_uist, nakano2010estimating, ishii2013gaze}. Among the limited works that do address missing data, classical interpolation techniques have been the primary method of choice: linear interpolation \cite{bulling2008s, huang2019saccalib}, nearest neighbours interpolation \cite{zheng2022identification}, and unsupervised Expectation-Maximization algorithm \cite{li2020selection}.
However, interpolation methods often struggle to accurately replicate the velocity distribution of real gaze data \cite{grootjen2023highlighting}. 
On the other hand, machine learning techniques have also been applied to gaze super-resolution - a special type of gaze imputation. Jiao et al. \cite{jiao23_uist} introduced SUPREYES, an implicit neural representation learning method for gaze super-resolution. However, it is important to note that SUPREYES primarily operates on low-resolution gaze data without any missing values, focusing on resolution enhancement rather than imputing missing data at arbitrary positions.
In stark contrast to previous methods, \methodName can impute missing gaze data at any arbitrary location while preserving the natural dynamics of human gaze behaviour by closely mimicking the gaze velocity distribution.

\subsection{Eye-head Coordination}

Eye-head coordination refers to the coordinated movements between the eyes and the head and has been extensively investigated in the areas of cognitive science and human-centered computing.
Specifically, Stahl studied the process of gaze shift and found that the amplitude of head movement is proportional to the amplitude of gaze shift~\cite{stahl1999amplitude}.
Hu et al. analysed eye-head coordination in immersive virtual environments and discovered that human eye gaze positions are strongly correlated with head rotation velocities~\cite{hu2019sgaze, hu2020dgaze, hu25hoigaze}.
Sidenmark et al. focused on the process of gaze shift in virtual reality and identified the coordination of eye, head, and body movements~\cite{sidenmark2019eye}.
Recently, inspired by the strong link between eye and head movements, researchers started to use both eye and head information in many applications and have achieved great success.
For example, Sidenmark et al.~\cite{sidenmark2019selection} employed users' eye and head movements in virtual reality to improve the accuracy of target selection.
Hu et al. predicted eye fixations in the future using historical gaze positions and head rotation velocities~\cite{hu2021fixationnet} and proposed to recognise the task a user is performing from user's eye and head movements~\cite{hu2022ehtask}.

Despite the fact that both eye and head motions are beneficial for many applications, they have not been studied together for eye gaze imputation yet. Existing future gaze prediction methods \cite{hu2019sgaze, hu2021fixationnet} cannot be directly applied to gaze imputation, since imputation requires bi-directional predictions. To the best of our knowledge, we are the first to demonstrate that eye-head coordination can be successfully transferred to the task of gaze imputation and lead to significant performance gains.

\subsection{Eye–hand Coordination}
Eye–hand coordination refers to the ability to synchronise visual input with hand movements, enabling precise motor control guided by visual feedback~\cite{johansson2001eyehand, land1999roles}. 
This coordination plays a crucial role in both human perception and action and has been widely leveraged to design interactive techniques in XR environments~\cite{ren2024eyehandapp, wang2024eyehandapp}. 
Beyond interaction design, eye–hand coordination has also been explored in gaze prediction. 
For example, Hu et~al.~\cite{hu24pose2gaze} analysed multiple datasets and demonstrated strong correlations between gaze directions and wrist movements across everyday activities, introducing Pose2Gaze, a model that predicts gaze direction from full-body poses. 
More recently, Hu et~al.~\cite{hu25hoigaze} proposed HOIGaze, which leverages two-hand gestures and 3D object information to improve gaze prediction performance during hand–object interactions. 

In stark contrast to these prior works that rely on motion capture systems for body or hand pose acquisition, \methodName{} utilises head movements from head-mounted devices and motion data from off-the-shelf wearable devices such as smartwatches and wristbands. 
\methodName{} achieves clear performance gains over previous state-of-the-art methods that depend on motion capture, demonstrating that meaningful eye–hand-head coordination can be effectively modelled using readily available consumer hardware.

\subsection{Time-Series Imputation}  
In theory, existing time-series imputation methods can be directly applied to gaze data imputation.  
Deep learning-based methods have been shown to outperform statistical approaches in prior time-series imputation studies. Existing methods explore various neural network architectures, including Convolutional Neural Networks (CNNs) \cite{wu2023timesnet}, Recurrent Neural Networks (RNNs) \cite{cao2018brits}, Transformers \cite{liu2024itransformer, zhang2023crossformer}, and Multilayer Perceptrons (MLPs) \cite{Zeng2022AreTE}.  
Generative methods such as Variational Autoencoders (VAEs), Generative Adversarial Networks (GANs), and diffusion models have also been explored for time-series imputation, including US-GAN \cite{miao2021generative}, GP-VAE \cite{fortuin2020gp}, and CSDI \cite{tashiro2021csdi}.  

While these methods offer powerful general-purpose solutions, they are typically designed for single-modal signals and do not account for the unique characteristics of gaze data or the availability of complementary head movement signals in mobile eye tracking. \change{In contrast, \methodName{} is specifically tailored to the gaze imputation task and is, to our knowledge, the first to explicitly incorporate head movement information, with an optional extension to wrist/hand movements captured by wearable devices. Our approach leverages eye-head coordination as its primary cue and, when wrist/hand signals are available, can further model eye-hand-head coordination to enhance gaze imputation performance.}

\section{Head-Assisted conditional diffusion model for Gaze  Imputation} 

\begin{figure*}[t]
    \centering
    \includegraphics[width=0.7\textwidth]{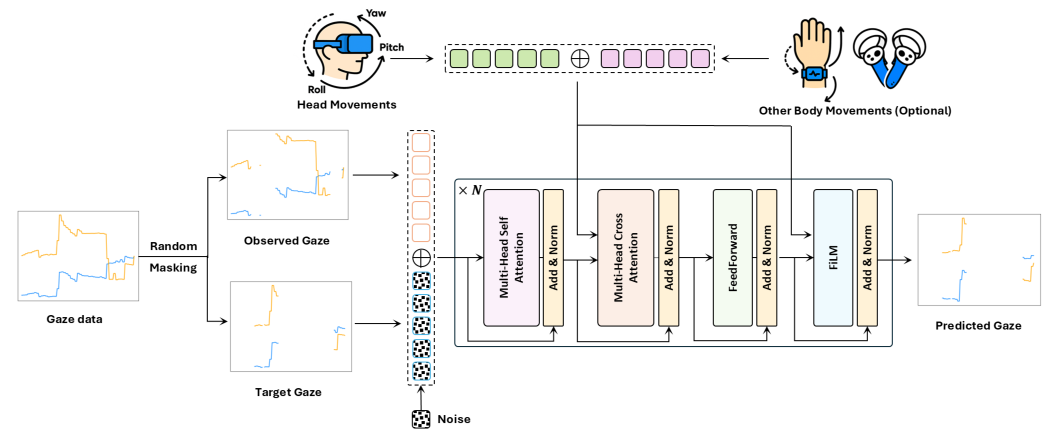}
    \caption{Overview of the training pipeline and model architecture of \methodName{}. 
  }
    \label{fig:training_pipeline}
\end{figure*}

Previous studies have demonstrated that human eye movements are strongly correlated with head movements, a phenomenon known as eye-head coordination \cite{sidenmark2019eye, sidenmark2019selection, stahl1999amplitude, hu2022ehtask, hu2019sgaze}.
Most commercially available head-mounted devices equipped with eye-tracking capabilities
are readily fitted with sensors for tracking users' head movements. 
Unlike gaze data, which inevitably contains missing values \cite{nystrom2025fundamentals, grootjen2024uncovering}, head movements are recorded continuously without any missing data, provided the sensor remains operational.
The key idea of \methodName is to exploit the close coordination between head and eye movements to impute missing values in gaze data.
At its core, \methodName employs a transformer-based diffusion model conditioned on observed gaze data and time-aligned head motion, enabling accurate and realistic gaze imputation .
Beyond head signals, the framework is designed to incorporate additional wrist movements, captured from commodity wearable devices such as smartwatches, wristbands, or XR hand controllers.
This extension exploits natural eye-hand coordination \cite{johansson2001eyehand, binsted2001eyehand} to further enhance imputation performance (see Section \ref{sec:diffusion} and \ref{sec:architecture}).
\change{Moreover, \methodName generalises to the extreme case of gaze generation (i.e., 100\% missing data), where it synthesises realistic gaze trajectories conditioned on head motion and, when available, wrist/hand motion captured by wearable sensors.}

\subsection{Problem Definition and Data Preparation} \label{sec:new_definition}
\paragraph{Head-assisted gaze imputation} 
For each gaze sequence $\mathbf{X}= \left\{x_1, x_2, ..., x_L \right\}$, we have a corresponding sequence of time-aligned head movements $\mathbf{H}= \left\{h_1, h_2, ..., h_L \right\}$ captured by the head-mounted eye trackers.
The task is to predict the gaze directions for missing values within $\mathbf{X}$ by leveraging both the head movements $\mathbf{H}$ and the existing gaze observations in $\mathbf{X}$.
Since head movements correspond to the movements of head-mounted devices, we obtain $\mathbf{H}$ by processing the SLAM poses of the head-mounted device.
Let us denote $T^{l}_{\text{world, tracker}} \in SE(3)$ as the pose of the mobile eye tracker in the world coordinate system at time step $l$.
The head movement at time step $l$ is represented as the relative transformation matrix between the SLAM poses at two consecutive time steps:
\begin{equation}
    h_l = \Delta T^{l, l+1}_{\text{tracker}} = (T^{l}_{\text{world, tracker}})^{-1}T^{l+1}_{\text{world, tracker}}
\end{equation}

For each gaze sample $x_l$, represented as $(\text{pitch}, \text{yaw})$, we normalise the values to the range $[0, 1]$ using a sine transformation before further processing.

\paragraph{Extension to wrist/hand movements.}  
When additional time-aligned signals from wrist-worn or hand-held devices (e.g., smartwatches, wristbands, or XR controllers) are available, the definition naturally extends to include an auxiliary modality $\mathbf{W} = \{w_1, w_2, \ldots, w_L\}$ of wrist/hand movements.  
Since both gaze and head are represented in the tracker-centric coordinate system, we express wrist motion at time step $l$ relative to the eye tracker.  
For example, consider a smart wristband worn on the left hand, with pose $T^{l}_{\text{world, band}} \in SE(3)$ at time step $l$.  
The relative transformation is then defined as:
\begin{equation}
    w_l = T^{l}_{\text{tracker, band}} 
        = \left(T^{l}_{\text{world, tracker}}\right)^{-1} T^{l}_{\text{world, band}}.
\end{equation}
This formulation allows \methodName to exploit eye--head--hand coordination for gaze imputation.

\subsection{\methodName Diffusion Process} \label{sec:diffusion}
At the core of \methodName lies a conditional diffusion model that performs gaze imputation by exploiting the strong correlations between eye and head movements, and optionally wrist movements when available.  
Although diffusion models have recently demonstrated strong performance in general time-series imputation tasks \cite{tashiro2021csdi}, they have not yet been adapted to handle multi-modal signals that exhibit natural eye--head coordination, such as the coupled dynamics of gaze and head motion.

To this end, \change{we design a motion-conditioned diffusion framework that integrates head movements and, in our wrist/hand extension, wrist movements throughout the denoising process.} These movements can be readily captured using commodity wearable devices, enabling practical deployment in real-world scenarios.
Let $\mathbf{x}_0$ denote the input gaze sequence, with $\mathbf{x}^{co}_0$ and $\mathbf{x}^{ta}_0$ representing the observed and target parts, respectively, separated by a randomly generated binary observation mask. The forward diffusion process in \methodName follows the same formulation as CSDI~\cite{tashiro2021csdi}.  
Given $\mathbf{B}$ as \change{the sequence of head movements and any available wrist/hand movements captured by wearable sensors,} corresponding to $\mathbf{x}_0$, the reverse process of \methodName is formulated as follows:

\begin{equation} \label{eq:sampling}
\resizebox{\columnwidth}{!}{$
p_\theta\left(\mathbf{x}^{ta}_{t-1} \mid \mathbf{x}^{ta}_t, \mathbf{x}^{co}_0, \mathbf{B}\right)=
    \mathcal{N}\left(\mathbf{x}^{ta}_{t-1} ; 
    \mu_\theta\left(\mathbf{x}^{ta}_t, t \mid \mathbf{x}^{co}_0,\mathbf{B}\right),
    \sigma_\theta\left(\mathbf{x}^{ta}_t, t \mid \mathbf{x}^{co}_0, \mathbf{B}\right) I\right)
$}
\end{equation}

As with standard denoising diffusion models \cite{ho2020denoising, tashiro2021csdi, zhang2024dismouse, shi2024actiondiffusion, kong2021diffwave, hu25haheae}, the training objective of \methodName is to accurately predict the added noise at diffusion time step $t$:
\begin{equation}
    \mathcal{L} = {\left\|{\epsilon_t - {\epsilon}_\theta\left(\mathbf{x}^{ta}_t, t \mid \mathbf{x}^{co}_0, \mathbf{B} \right)}\right\|}_2 
\end{equation}

Figure \ref{fig:training_pipeline} shows the \methodName training pipeline. The summarisations of \methodName's training and sampling procedures can be found in the supplementary.

\subsection{\methodName Architecture} \label{sec:architecture}
Figure \ref{fig:training_pipeline} shows an overview of the \methodName architecture.
The primary challenge lies in effectively integrating information from \change{head movements and optional wrist/hand movements} captured by commodity wearable devices with the observed gaze sequence $\mathbf{x}^{co}_0$ to accurately reconstruct the missing gaze samples $\mathbf{x}^{ta}_0$.  
\change{To this end, we design a transformer-based diffusion model that captures cross-modal dependencies between eye and head representations and can optionally incorporate wrist motions to exploit gaze--wrist coordination in settings where such signals are available.}
We additionally include a FiLM-based skip-fusion mechanism that re-injects the motion context at each Transformer block.
This design enables \methodName to learn a more coherent correlation between head and gaze dynamics, leading to more accurate and realistic gaze imputation.

\vspace{.1cm}
\textbf{Input mapping.} As outlined in Sections \ref{sec:new_definition}, the input gaze data is represented as a sequence of $\text{(pitch, yaw)}$ with shape $(L, 2)$, where $L$ denotes the sequence length.
The input gaze data is then divided into the observed gaze $\mathbf{x}^{co}_0 \in \mathbb{R}^{L \times 2}$ and the noisy target part $\mathbf{x}^{ta}_t \in \mathbb{R}^{L \times 2}$ using a randomly generated mask and noise addition.

The head movements captured by the mobile eye tracker are denoted as $\mathbf{H} \in \mathbb{R}^{L \times 4 \times 3}$, consisting of a sequence of transformations $h_l = [R, \mathbf{t}] \in \mathbb{R}^{4 \times 3}$, where $R \in \mathbb{R}^{3 \times 3}$ represents head rotation and $\mathbf{t} \in \mathbb{R}^{1 \times 3}$ corresponds to head translation. 
Following established practice for processing transformation matrices~\cite{yi2024egoallo}, each transformation is first flattened into a 12-dimensional vector, after which we apply Fourier positional encoding to obtain the encoded head movements $\mathbf{H} \in \mathbb{R}^{L \times 12}$.  

When \change{additional wrist or hand movements} $\mathbf{W} \in \mathbb{R}^{L \times 4 \times 3}$ are available, we apply the same Fourier encoding to $\mathbf{W}$.  

For each input modality, including the binary observation mask $\mathbf{M}$, we project the features into a shared latent space of dimension $D = 64$ using a multilayer perceptron (MLP) followed by a GELU activation. 
To preserve temporal order, sinusoidal positional encodings~\cite{vaswani2017attention} of shape $(L, D)$ are added to each projected sequence, except for the observation mask. 
The noisy and observed gaze features are then concatenated along the feature dimension to form a unified gaze tensor $\mathbf{G} \in \mathbb{R}^{L \times 2D}$. 
Finally, the projected head features, conditional mask, and—when available—\change{wrist/hand-motion features} are concatenated along the feature dimension to form a unified motion context tensor $\mathbf{B} \in \mathbb{R}^{L \times D_\text{motion}}$, where $D_\text{motion} = N_\text{motion} \times D$ and $N_\text{motion}$ denotes the number of conditional modalities. 
This motion context serves as auxiliary input to guide the gaze imputation process.

\vspace{.1cm}
\textbf{Transformer block.} 
At the core of the \methodName architecture lies a stack of $N$ transformer blocks. 
In each block, we first map the diffusion step $t$ to the same dimension as the gaze tensor $\mathbf{G}$ and merge it via element-wise addition, enabling the model to adapt its denoising behaviour according to the current diffusion timestep.

\paragraph{Within-gaze correlation.} 
The gaze features (of dimension $2D$) undergo multi-head self-attention to capture temporal dependencies within the gaze sequence. Layer normalisation is applied before attention, and the output is added to the input via a residual connection.

\paragraph{Cross-modality correlation.} 
Since prior work has shown that cross-attention effectively captures relationships between different modalities~\cite{hu25hoigaze}, we introduce a multi-head cross-attention mechanism that attends from the gaze features (queries of dimension $2D$) to the motion context tensor (keys and values of dimension $D_{motion}$). The query, key, and value tensors are independently normalised before attention computation using separate layer normalisation layers.
This mechanism allows the model to dynamically attend to the most relevant motion cues at each timestep, thereby fusing multimodal information for more accurate gaze imputation. 

Similar to a standard Transformer~\cite{vaswani2017attention}, the attended features are further refined by a feed-forward network (FFN) with GELU activation. 

\paragraph{Hybrid fusion mechanism.} 
Information from the motion context can become attenuated as it propagates through successive linear and nonlinear transformations within the feed-forward network.
To preserve this conditioning signal, inspired by the skip connections in ResNet~\cite{he2016deep}, we add a \textit{skip fusion} mechanism that re-injects head and optional wrist/hand information at each Transformer block via feature-wise linear modulation (FiLM)~\cite{perez2018film}.
This module provides an additional path for motion conditioning.

Specifically, after input mapping, the motion features (head and wrists) are mean-pooled over the temporal dimension and concatenated, then projected to dimension $D$ to form a global context vector $\mathbf{C} \in \mathbb{R}^D$. This context is passed through a learned projection to produce scale and shift parameters $\gamma(\mathbf{C}), \beta(\mathbf{C}) \in \mathbb{R}^{2D}$. These parameters modulate the layer-normalised features: $\text{FiLM}(\mathbf{K}, \mathbf{C}) = \text{LN}(\mathbf{K}) \odot (1 + \gamma(\mathbf{C})) + \beta(\mathbf{C})$, where $\mathbf{K}$ is the enhanced feature after the FFN. The modulated features are then passed through an output projection layer and added to the output of the FFN.

\vspace{.1cm}
\textbf{Output projection.} 
The output from the final Transformer block is projected through an MLP to predict the added noise estimate, which is used in the reverse diffusion process to iteratively reconstruct the imputed gaze sequence.

\subsection{Gaze Generation} \label{sec: method_generation}

\change{Gaze generation can be seen as a special case of gaze imputation, where the missing ratio reaches 100\%. 
In this setting, no gaze observations are available, and the model must synthesise an entire gaze trajectory from auxiliary motion signals alone.}
By removing the observed gaze input and the corresponding conditional mask, \methodName naturally functions as a conditional generative model that produces realistic gaze sequences based solely on head movements and, when available, \change{wrist/hand-motion signals.}

\section{Experiments - Head-assisted Gaze Imputation}\label{sec: exp}
We first conducted experiments to evaluate the performance of \methodName on head-assisted gaze imputation. 

\subsection{Datasets}
We evaluate \methodName{} on three large-scale egocentric datasets containing mobile eye tracking and head motion during everyday activities.

Nymeria~\cite{ma2024nymeria} is the largest in-the-wild human motion dataset, providing 30\,Hz gaze, head motion, wrist motion. We use it for training and testing, selecting recordings with well-calibrated personalised gaze and valid eye tracker poses. The data are split into 80\% training (593 recordings, 145.8\,h), 5\% validation (38 recordings, 8.4\,h), and 15\% test (111 recordings, 28.7\,h).

Ego-Exo4D~\cite{grauman2024ego} offers synchronised egocentric and exocentric recordings of skilled activities. Its egocentric subset includes 30\,Hz gaze and SLAM-derived head poses. We use all 72 well-calibrated recordings (4.6\,h) for cross-dataset evaluation.

HOT3D ~\cite{banerjee2024hot3d} focuses on hand–object interactions in indoor environments. We use all 111 well-calibrated egocentric recordings (3.48\,h) with gaze and SLAM poses for cross-dataset evaluation.

\subsection{Evaluation Settings}

\paragraph{Baselines}  
We compared \methodName against several baseline approaches.
These include Head direction as proxy \cite{hu2020dgaze, hu24pose2gaze}, widely used interpolation methods for gaze imputation, including Linear interpolation and Nearest interpolation, as well as state-of-the-art deep learning-based time-series imputation techniques: iTransformer \cite{liu2024itransformer}, DLinear \cite{Zeng2022AreTE}, TimesNet \cite{wu2023timesnet}, BRITS \cite{cao2018brits} and CSDI \cite{tashiro2021csdi}. \change{In addition, we compared with Pose2Gaze \cite{hu24pose2gaze}, the state-of-the-art gaze generation method that predicts the full gaze sequence without access to observed gaze from full body poses. Its performance is evaluated on the same missing frames as other imputation models. }

Moreover, we compare \methodName with its predecessor, \methodNameOld~ \cite{jiao25_uist}, to highlight the improvements introduced by the newly proposed architecture.

\paragraph{Input duration.}
We chose five seconds as the duration for all inputs, same as prior gaze-based deep learning models \cite{9865991, jiao2024diffeyesyn}. 
We clipped the gaze recordings in the Nymeria dataset into non-overlapped five-second segments according to the time stamps in the atomic motion descriptions.
For the Ego-Exo4D and HOT3D datasets, we ignored the data within the very first and last second in each recording to ensure the data quality and clipped the recording into non-overlapping five-second segments.
Following prior work \cite{9865991}, all segments with more than 5\% invalid gaze samples were discarded. To ensure the model only learns to reconstruct real gaze movements, all the remaining invalid gaze samples were excluded from the loss computation during training.

\paragraph{Data loss ratio.}  
The selected data loss ratios were informed by blink duration statistics and missing data ratios reported in prior research:
Blinks occur approximately 20 times per minute, with each blink lasting between 150–450 milliseconds \cite{stern1984endogenous, nystrom2024blink}. 
Consequently, up to 10\% missing data can be attributed solely to blinks \cite{nystrom2025fundamentals}. 
Additionally, prior studies have reported missing data ratios ranging from 20\% to 60\% \cite{holmqvist2012eye, schnipke2000trials, pernice2009conduct}. Moreover, an empirical research shows that consecutive missing values last 1,325 milliseconds on average with the standard deviation of 4,076 milliseconds, and the majority of the missing segments are shorter than 1 second \cite{grootjen2024investigating}.
Based on these findings, we evaluated \methodName under missing data conditions of 10\%, 30\%, and 50\%.
Furthermore, to assess \methodName's robustness in extreme scenarios, we also tested it with 90\% missing data.
Since the longest blink duration (450 ms) corresponds to approximately 10\% of our five-second input window, we primarily evaluated \methodName on long blinks using a 10\% missing ratio.
Specifically, for each gaze trajectory, we randomly masked a continuous 10\% segment of valid data.
For 30\% and 50\% missing ratios, we simulated real-world data loss by ensuring that each masked segment lasted at least 150 milliseconds, corresponding to the shortest blink duration. For the 90\% missing ratio, we simulated an extreme case in which 90\% of the sequence was continuously missing, reflecting real-world scenarios involving prolonged gaze loss as reported by~\cite{grootjen2024investigating}.

\paragraph{Evaluation metrics.} We used two metrics for evaluation:
\begin{itemize}[leftmargin = *]
\item  Mean Angular Error (MAE): MAE is the most commonly used evaluation metric in previous gaze estimation research \cite{hu2020dgaze, hu2019sgaze, hu24pose2gaze}. It measures the angular difference (in degrees) between the predicted and the ground truth gaze vectors. 

\item Jensen–Shannon divergence (JS): While MAE evaluates the accuracy of gaze direction reconstruction, it does not assess whether the imputed gaze movements exhibit realistic human gaze dynamics. To address this, we incorporated JS divergence to measure the velocity distribution between generated data and human data.
JS divergence ranges between zero and one, with lower values indicating better performance.
\end{itemize}
A good gaze imputation method should achieve lower JS on the basis of lower MAE.

\paragraph{Implementation details.} All datasets used in our experiments provide gaze data at 30 Hz, and we set the input duration to five seconds, resulting in a total sequence length of $L = 150$.
For \methodName, we set the number of transformer blocks to $N=4$ and the number of attention heads to eight for self-attention and cross-attention mechanism. The diffusion process consisted of $T=50$ steps, using a cosine noise schedule with a minimum noise level of $1-\alpha_{1} = 10^{-4}$ and a maximum noise level of $1-\alpha_{T} = 0.5$. The training was conducted for 500 epochs with a batch size of 256, using the AdamW optimizer with an initial learning rate of $10^{-3}$, which decayed to $10^{-4}$ at epoch 375 and further to $10^{-5}$ at epoch 450.

All deep learning-based methods were trained on the Nymeria training set, and we selected the best-performing model on the validation set for the final evaluation. 
\change{For Pose2Gaze, we adapted the input body poses to the 23-joint format provided in the Nymeria.} 
For generative models that do not produce deterministic outputs, we followed \cite{tashiro2021csdi} and used the median of 100 generated samples for evaluation.

\subsection{Gaze Imputation Results} \label{sec:results}

\paragraph{Quantitative results.}

\begin{table}[]
\centering
\begin{tabular}{lcrrr}
\toprule
Nymeria        & \multicolumn{4}{c}{Data loss ratio}                           \\ \midrule
               & 10\%           & 30\%          & 50\%          & 90\%          \\ \midrule
Head direction & 23.32         & 23.43         & 23.44         & 23.44         \\ \midrule
\change{Pose2Gaze \cite{hu24pose2gaze}} & \change{13.01}         & \change{13.05}         & \change{13.05}         & \change{13.07}         \\ \midrule
Linear         & 4.96          & 6.88          & 9.68          & 11.54         \\
Nearest        & 5.29          & 6.52          & 8.34          & 12.61         \\ \midrule
iTransformer \cite{liu2024itransformer}   & 8.75          & 11.10         & 16.57         & 24.05         \\
DLinear  \cite{Zeng2022AreTE}      & 12.12         & 12.25         & 12.97         & 14.01         \\
TimesNet \cite{wu2023timesnet}      & 19.53         & 18.59         & 20.32         & 22.91         \\
BRITS  \cite{cao2018brits}        & 10.25         & 11.98         & 14.06         & 17.58         \\
CSDI \cite{tashiro2021csdi}          &  4.72    &  5.90    &  7.44    &  10.54   \\ \midrule
\methodNameOld            & {\ul 3.67} & {\ul 4.55} & {\ul 5.77} & {\ul 8.53} \\ 
\methodName            & \textbf{3.54} & \textbf{4.40} & \textbf{5.58} & \textbf{8.18} \\ \bottomrule
\end{tabular}
\caption{MAE of gaze imputation across different methods and data loss ratios on the Nymeria \cite{ma2024nymeria} test set. The best results are marked in bold, and the second-best are underlined.} \label{tab: mae_nymeria}
\end{table}

\begin{table}[]
\centering
\begin{tabular}{lrrrr}
\toprule
Nymeria        & \multicolumn{4}{c}{Data loss ratio}                                         \\ \midrule
               & \multicolumn{1}{c}{10\%} & 30\%           & 50\%           & 90\%           \\ \midrule
Head direction & 0.139                    & 0.137          & 0.139          & 0.146          \\ \midrule
Linear         & 0.129                    & 0.078          & 0.089          & 0.150          \\
Nearest        & 0.081                    & 0.073          & 0.103          & 0.135          \\ \midrule
CSDI \cite{tashiro2021csdi}           & {\ul 0.044}              & {\ul 0.042}    & 0.037    & {\ul 0.030}    \\ \midrule
\methodNameOld            & \textbf{0.042}           & \textbf{0.040} & \textbf{0.035} & \textbf{0.017} \\ 
\methodName            & 0.045           & {\ul 0.042} & {\ul 0.036} & \textbf{0.017} \\ \bottomrule
\end{tabular}
\caption{JS of gaze imputation across different methods and data loss ratios on the Nymeria \cite{ma2024nymeria} test set. The best results are marked in bold, and the second-best are underlined.} \label{tab: js_nymeria}
\end{table}

Table~\ref{tab: mae_nymeria} presents the MAE of gaze imputation across different methods and missing ratios for within-dataset evaluation on the Nymeria test set. As shown in the table, \methodName consistently outperforms all baselines across all levels of missing data.
Furthermore, the additional on average 3.6\% improvement over \methodNameOld across all missing ratios demonstrates that the new model architecture more effectively captures eye--head coordination.
\change{To verify that these improvements are statistically meaningful, we conducted paired Wilcoxon signed-rank tests on per-sequence angular errors. The performance gains of \methodName{} over both \methodNameOld{} and CSDI were statistically significant across all missing ratios ($p < 0.001$).}

In contrast, except for CSDI, time-series imputation methods did not achieve performance comparable to traditional interpolation methods.
This suggests that these time-series imputation methods cannot be directly adapted to gaze data without modification.
\change{Similarly, because Pose2Gaze cannot see any observed gaze context, it also performed worse than traditional interpolations.} Consequently, we excluded these methods from further evaluations (see our supplementary materials if interested).

Table \ref{tab: js_nymeria} shows the JS of gaze imputation across different methods and data loss ratios, also on the Nymeria test set.
The velocity distributions of gaze imputed by \methodName achieve performance on par with \methodNameOld across all experimental settings. This demonstrates that \methodName preserves the natural dynamics of human gaze while simultaneously achieving the lowest MAE, offering both higher accuracy and greater biological plausibility.
In contrast, traditional interpolation methods perform substantially worse in replicating human-like eye movements than diffusion-based approaches.

\begin{table*}[t]
\resizebox{\textwidth}{!}{
\begin{tabular}{lrrrrrrrrrrrrrrrr}
\toprule
                & \multicolumn{8}{c}{Mean angular error (MAE)}                                                                                                                                                 & \multicolumn{8}{c}{Jensen–Shannon divergence (JS)}                                                                                                                                                         \\ \midrule
Dataset         & \multicolumn{4}{c}{Ego-Exo4D}                                                                & \multicolumn{4}{c}{HOT3D}                                               & \multicolumn{4}{c}{Ego-Exo4D}                                                                   & \multicolumn{4}{c}{HOT3D}                                                   \\
Data loss ratio & \multicolumn{1}{c}{10\%} & 30\%          & 50\%          & \multicolumn{1}{r}{90\%}          & \multicolumn{1}{c}{10\%} & 30\%          & 50\%          & 90\%          & \multicolumn{1}{c}{10\%} & 30\%           & 50\%           & \multicolumn{1}{r}{90\%}           & \multicolumn{1}{c}{10\%} & 30\%           & 50\%           & 90\%           \\ \midrule
Head direction  & 25.82                    & 25.76         & 25.88         & \multicolumn{1}{r}{25.82}         & 23.63                    & 23.81         & 23.70         & 23.73         & 0.126                    & 0.125          & 0.124          & \multicolumn{1}{r}{0.131}          & 0.148                    & 0.148          & 0.147          & 0.145          \\
Linear          & 3.92                     & 5.54          & 7.90          & \multicolumn{1}{r}{9.29}          & 3.87                     & 5.64          & 7.80          & 8.98          & 0.094                    & 0.085          & 0.073          & \multicolumn{1}{r}{0.135}          & 0.277                    & 0.102          & 0.096          & 0.148          \\
Nearest         & 4.18                     & 5.18          & 6.68          & \multicolumn{1}{r}{10.13}         & 4.14                     & 5.23          & 6.61          & 9.86          & 0.062                    & { 0.066}    & 0.081          & \multicolumn{1}{r}{0.121}          & 0.080                    & 0.081          & 0.100          & 0.135          \\
CSDI  \cite{tashiro2021csdi}          & { 3.78}               & { 4.78}    & { 6.07}    & \multicolumn{1}{r}{{ 8.91}}    & { 3.67}               & { 4.69}    & { 5.85}    & { 8.28}    & {\ul 0.042}              & \textbf{0.041} & \textbf{ 0.033}    & \multicolumn{1}{r}{{ 0.026}}    & {\ul 0.052}              & {\ul 0.051}    & {\ul 0.042}    & { 0.029}    \\ \midrule
\methodNameOld            & {\ul 3.06}            & {\ul 3.80} & {\ul 4.86} & \multicolumn{1}{r}{{\ul 7.37}} & {\ul 3.01}            & {\ul 3.91} & {\ul 4.99} & {\ul 7.34} & \textbf{0.040}           & \textbf{0.041} & \textbf{0.033} & \multicolumn{1}{r}{{\ul 0.024}} & \textbf{0.049}           & \textbf{0.050} & \textbf{0.040} & \textbf{0.021} \\
\methodName            & \textbf{2.98}            & \textbf{3.73} & \textbf{4.76} & \multicolumn{1}{r}{\textbf{7.08}} & \textbf{2.94}            & \textbf{3.84} & \textbf{4.86} & \textbf{7.19} & 0.044           & {\ul 0.043} & {\ul 0.034} & \multicolumn{1}{r}{\textbf{0.023}} & 0.055           & 0.053 & {\ul 0.042} & {\ul 0.022} \\ \bottomrule
\end{tabular}}
\caption{Cross-dataset evaluation results: MAE and JS of gaze imputation across different methods and data loss ratios on the Ego-Exo4D \cite{grauman2024ego} and HOT3D \cite{banerjee2024hot3d} datasets. The best results are marked in bold, and the second-best are underlined.} \label{tab: cross_dataset}
\end{table*}

We then conducted a cross-dataset evaluation on the Ego-Exo4D and HOT3D datasets to assess whether \methodName{} can generalise to diverse everyday settings. 
The results are shown in Table \ref{tab: cross_dataset}. 
Similar to the results on the Nymeria dataset, despite \methodName{} not being trained or fine-tuned on these datasets, it achieved the lowest MAE across all methods and missing ratios.
Moreover, the additional 2.4\% average improvement over \methodNameOld{} across all missing ratios demonstrates that the new model architecture more effectively captures eye--head coordination. \change{These improvements were also statistically significant according to paired Wilcoxon signed-rank tests conducted on the Ego-Exo4D and HOT3D datasets (all $p < 0.001$).} Finally, \methodName{} attained a comparable JS divergence to \methodNameOld{}, indicating that its imputed gaze velocity distribution remains close to the natural human gaze dynamics.  
Moreover, it is worth noting that in the cross-dataset evaluation, the absolute MAE across different missing ratios did not decrease compared with the within-dataset results on the Nymeria dataset (see Table \ref{tab: mae_nymeria}).
This suggests that \methodName{} learns robust correlations between eye and head movements that generalise well across datasets featuring diverse everyday activities and environments.

\paragraph{Qualitative results.} 
\begin{figure*}[t]
    \centering
    \includegraphics[width=1.5\columnwidth]{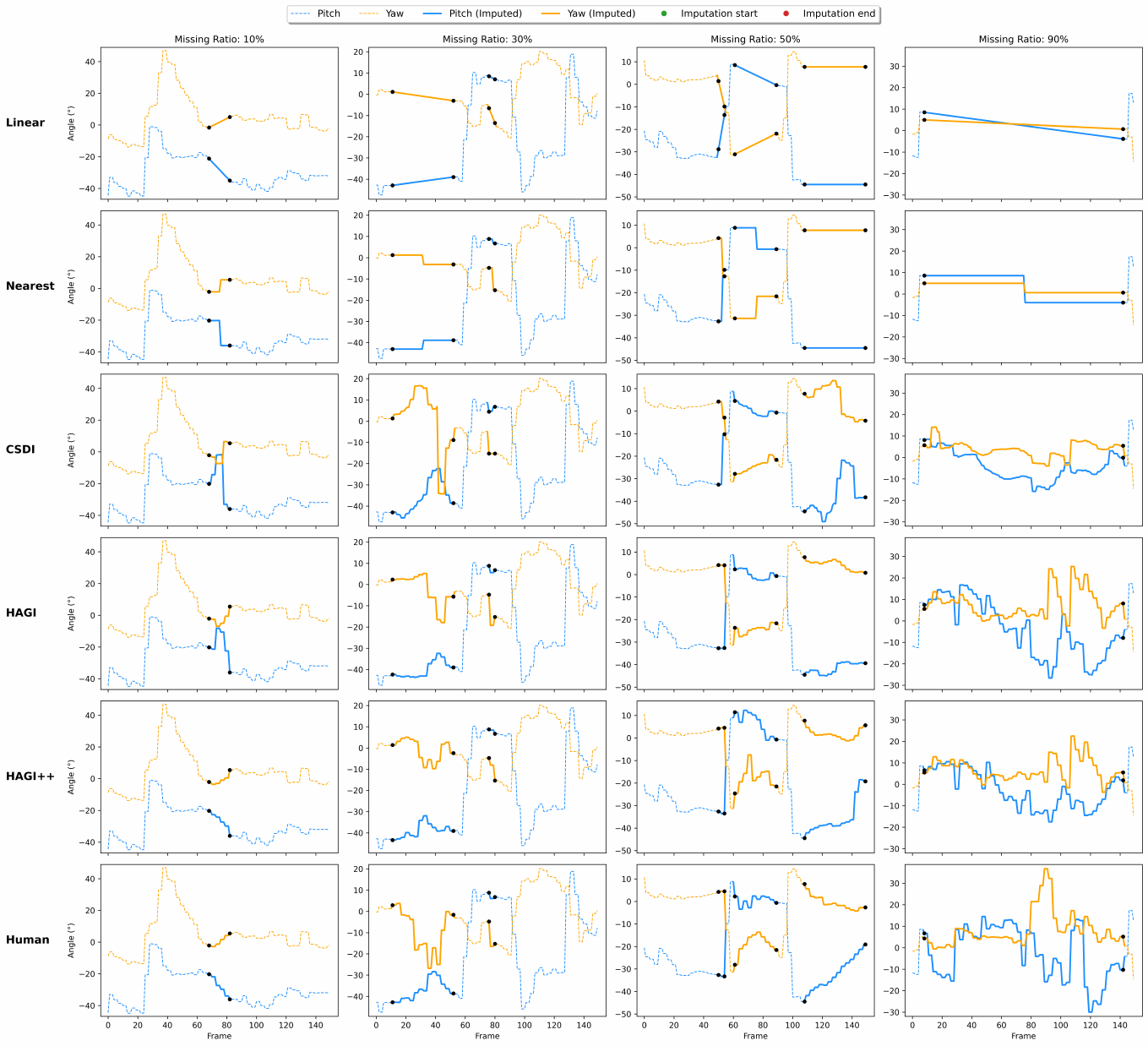}
    \caption{Four examples of gaze imputation results at different missing ratios (10\%, 30\%, 50\%, 90\%) using different methods in the cross-dataset evaluation.}  
    \label{fig:qual}
\end{figure*}


We present four sample gaze imputation results from the cross-dataset evaluation for five best-performing methods in Figure \ref{fig:qual}.
\methodName{} benefits from the incorporation of head movement information, producing imputed gaze trajectories that more closely follow the ground-truth human eye movements both spatially and temporally. 
Even in the challenging scenario with 90\% missing values, both \methodName{} generate gaze samples that remain well aligned with the ground truth.
This suggests that leveraging head movements provides valuable contextual information for gaze imputation, enabling \methodName{} to generate more naturalistic and human-like gaze trajectories than baselines, aligning with results of MAE in Table \ref{tab: cross_dataset}.
In contrast, the gaze trajectories imputed by traditional interpolation methods are visually dissimilar to real human eye movements.
This visual assessment aligns with the JS values reported in Table \ref{tab: cross_dataset}.
Although CSDI achieved a low JS score in Table \ref{tab: cross_dataset}, its imputed gaze data exhibits substantial spatial deviation from real human eye movements.

\subsection{Head Rotation vs. Translation.}\label{sec:head_ablation}
The results so far indicate that incorporating head information allows \methodName{} to outperform single‑modal baselines.
However, head movements contain both rotation and translation, and their individual contributions to gaze imputation remain unclear.
To investigate this, we trained two ablated variants of \methodNameOld and \methodName{}: one using only head rotation matrices and another using only head translation vectors.  

The results are shown in Table \ref{tab:head_ablation}.
Since prior work \cite{jiao2024diffgaze, jiao2024diffeyesyn} and our previous findings demonstrated that diffusion-based approaches effectively model gaze velocity distributions, we report only the MAE.
Compared with CSDI, which does not use head information, \methodName{} trained with only head rotation achieves lower MAE across all settings, and the translation‑only variant performs better in most cases. This indicates that both rotation and translation \change{provide useful predictive cues} and contribute to performance gains.
Moreover, head rotation consistently outperforms translation, suggesting it plays a stronger role in gaze imputation. The full \methodName{}, which uses the complete head transformation, surpasses both ablations, demonstrating the benefit of leveraging the full head motion. Notably, even the rotation‑only variant outperforms \methodNameOld{}, highlighting the advantages of the new architecture.

\begin{table*}[]
\centering
\begin{tabular}{lcccccccccccrrr}
\toprule
                  & \multicolumn{2}{c}{Head movements} & \multicolumn{4}{c}{Nymeria }                                  & \multicolumn{4}{c}{Ego-Exo4D }                                & \multicolumn{4}{c}{HOT3D }                                    \\
                  & Rotation        & Translation       & 10\%          & 30\%          & 50\%          & 90\%          & 10\%          & 30\%          & 50\%          & 90\%          & 10\%          & 30\%          & 50\%          & 90\%          \\ \midrule
CSDI  \cite{tashiro2021csdi}            & \XSolidBrush               & \XSolidBrush                & 4.72          & 5.90          & 7.44          & 10.54         & 3.78          & 4.78          & 6.07          & 8.91          & 3.67          & 4.69          & 5.85          & 8.28          \\ \midrule
\multirow{3}{*}{\methodNameOld} & \XSolidBrush               & \Checkmark                 & 4.66          & 5.75          & 7.29          & 10.47         & 3.72          & 4.63          & 5.91          & 8.79          & 3.60          & 4.57          & 5.74          & 8.19          \\
                  & \Checkmark               & \XSolidBrush                 & {4.41}& {5.36}          & {6.66 }         & {9.56}          & {3.55}          & {4.35}          & {5.43}          & {8.23}          & {3.42}          & {4.33}          & {5.38}          & {7.79}          \\
                  & \Checkmark              & \Checkmark                 & {3.67} & {4.55} & {5.77} & {8.53} & {3.06} & {3.80} & {4.86} & {7.37} & {3.01} & {3.91} & {4.99} & {7.34} \\ \midrule
\multirow{3}{*}{\methodName} & \XSolidBrush               & \Checkmark                 & 4.65          & 5.79          & 7.34          & 10.65         & 3.74          & 4.63          & 5.89          & 8.78          & 3.61          & 4.60          & 5.75          & 8.38          \\
                  & \Checkmark               & \XSolidBrush                 & {\ul3.60}& {\ul4.48}          & {\ul5.69 }         & {\ul 8.42}          & {\ul3.02}          & {\ul 3.79}          & {\ul 4.86}          & {\ul 7.36}          & {\ul 2.95}          & {\ul 3.88}          & {\ul 4.95}          & {\ul 7.36}          \\
                  & \Checkmark              & \Checkmark                 & \textbf{3.54} & \textbf{4.40} & \textbf{5.58} & \textbf{8.18} & \textbf{2.98} & \textbf{3.73} & \textbf{4.76} & \textbf{7.08} & \textbf{2.94} & \textbf{3.84} & \textbf{4.86} & \textbf{7.19} \\ \bottomrule
\end{tabular}
\caption{The MAE results of the ablation study on head rotation and translation across different data loss ratios in the Nymeria \cite{ma2024nymeria}, Ego-Exo4D \cite{grauman2024ego}, and HOT3D \cite{banerjee2024hot3d} datasets. The best results are marked in bold, and the second-best are underlined.}\label{tab:head_ablation}

\end{table*}

\subsection{\change{Head–Gaze Discrepancy vs. \methodName Performance}}

\begin{figure}[!h]
    \centering
    \includegraphics[width=\columnwidth]{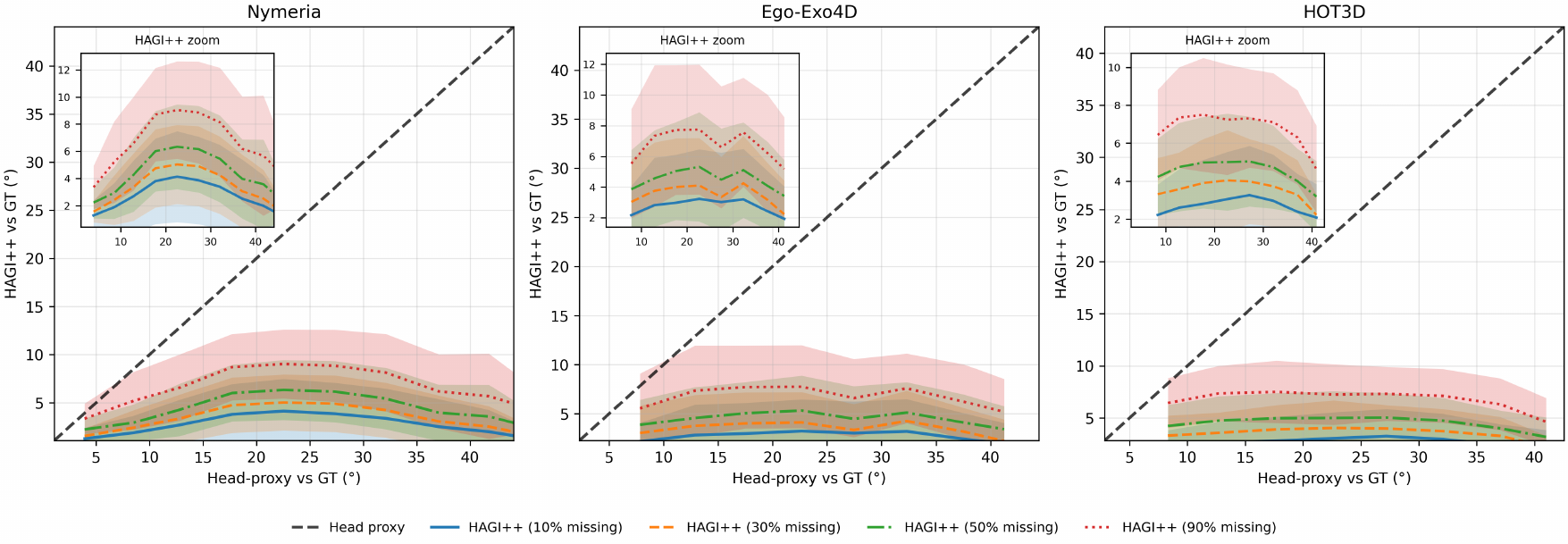}
    \caption{\change{Head–gaze discrepancy vs. \methodName prediction error. Each panel shows one dataset. x-axis: MAE between the head proxy and GT gaze (per sequence). y-axis: MAE between \methodName and GT gaze on masked frames. The dashed diagonal is the baseline where the head direction is a proxy of gaze. Colored curves show \methodName at four missing ratios; shaded regions are the standard deviation across sequences within each head-proxy bin. \methodName errors remain low and largely flat as head-proxy error increases, demonstrating that our predictions do not simply replicate head orientation.} }
    \label{fig:head_gaze_ana}
\end{figure}

\change{
A common concern in egocentric gaze modelling is whether gaze can be approximated by head orientation alone. To investigate this, we analyse the relationship between head--gaze discrepancy and HAGI++ prediction error on the three datasets.
For each test sequence, we compute the MAE between the ground-truth gaze and the head proxy, as well as the MAE between the ground-truth gaze and \methodName predictions on masked frames. Figure~\ref{fig:head_gaze_ana} plots \methodName prediction error against head-proxy error under different missing ratios.
\methodName prediction error remains relatively stable as head--gaze discrepancy increases, rather than following the head-proxy baseline. This indicates that HAGI++ does not simply learn a head-direction prior, but instead captures eye--head coordination to recover gaze behaviour even when gaze substantially deviates from head orientation.}

\subsection{Efficiency Comparison}
Since real‑time imputation during data collection is a potential application, we compared GPU memory usage and inference time for imputing a batch of 512 five‑second gaze sequences across the three best‑performing methods. All models were run with the same inference script on a single NVIDIA Tesla V100 (32 GB) GPU. As shown in Table \ref{tab: memory}, \methodName{} requires substantially less memory and achieves faster inference than the baselines, while also delivering the most accurate gaze predictions.

\begin{table}[h]
\centering
\begin{tabular}{lrr}
\toprule
       & Memory (MB) & Inference Time (s)                                         \\ \midrule
CSDI \cite{tashiro2021csdi}           & {\ul 3848}              & {\ul 4.895}    \\ 
\methodNameOld            & 4380          & 7.484  \\ 
\methodName            & \textbf{2608}           &  \textbf{3.455} \\ \bottomrule
\end{tabular}
\caption{The GPU memory consumption and inference time for three methods tested on one Nvidia Tesla V100 (32GB) GPU with a batch size of 512 during inference. The best results are marked in bold, and the second-best are underlined.} \label{tab: memory}
\end{table}

\subsection{Ablation Study}\label{sec:ablation}
\change{We conducted an ablation study to assess the contribution of the key components in \methodName{}. As shown in Table \ref{tab:method_ablation}, incorporating head information yields a substantial 19.6\% average improvement over the gaze-only variant. Adding the FiLM-based skip fusion further improves average MAE by 0.7\%. Even the gaze-only and head-only variants remain strong over baselines, underscoring the robustness of the underlying diffusion architecture and the importance of head-motion conditioning.}

\begin{table*}[]
\begin{tabular}{lccrrrrrrrrrrrr}
\toprule
                  & \multicolumn{2}{c}{\methodName components} & \multicolumn{4}{c}{Nymeria }                                                                               & \multicolumn{4}{c}{Ego-Exo4D }                                                                             & \multicolumn{4}{c}{HOT3D }                                                \\
                  & Head information  & FiLM fusion & \multicolumn{1}{c}{10\%} & \multicolumn{1}{c}{30\%} & \multicolumn{1}{c}{50\%} & \multicolumn{1}{c}{90\%} & \multicolumn{1}{c}{10\%} & \multicolumn{1}{c}{30\%} & \multicolumn{1}{c}{50\%} & \multicolumn{1}{c}{90\%} & \multicolumn{1}{c}{10\%} & 30\%          & 50\%          & 90\%          \\ \midrule
CSDI  \cite{tashiro2021csdi}            & \XSolidBrush               & \XSolidBrush                & 4.72          & 5.90          & 7.44          & 10.54         & 3.78          & 4.78          & 6.07          & 8.91          & 3.67          & 4.69          & 5.85          & 8.28          \\ \midrule
\methodNameOld   & \Checkmark             & \XSolidBrush            & {3.67}            & {4.55}            & {5.77}            & {8.53}             & {3.06}            & {3.80}            & {4.86}            & {7.37}             & {3.01}            & {3.91} & {4.99} & {7.34} \\ \midrule
\multirow{3}{*}{\methodName (Ours)}      & \XSolidBrush           & \XSolidBrush & 4.61                     & 5.74                     & 7.27                     & 10.38                      & 3.70                     & 4.61                     & 5.92                     & 8.78                      & 3.58                     & 4.56          & 5.72          & 8.23          \\
                              &  \Checkmark          & \XSolidBrush & {\ul 3.55}                     & {\ul 4.42}                    & {\ul 5.61}                    & {\ul 8.24}                      & {\ul 2.99}                     & {\ul 3.75}                     & {\ul 4.79}                     & {\ul 7.18}                      & {\ul 2.96 }                    & {\ul 3.86}          & {\ul 4.90}          & {\ul 7.26}          \\
                  & \Checkmark                        & \Checkmark & \textbf{3.54}            & \textbf{4.40}            & \textbf{5.58}            & \textbf{8.18}             & \textbf{2.98}            & \textbf{3.73}            & \textbf{4.76}            & \textbf{7.08}             & \textbf{2.94}            & \textbf{3.84} & \textbf{4.86} & \textbf{7.19} \\ \bottomrule
\end{tabular}
\caption{The ablation study results (MAE) on different \methodName components across different data loss ratios in the Nymeria \cite{ma2024nymeria}, Ego-Exo4D \cite{grauman2024ego}, and HOT3D \cite{banerjee2024hot3d} datasets. The best results are marked in bold, and the second-best are underlined.}\label{tab:method_ablation}
\end{table*}

\section{Experiments – Head–Wrist-Assisted Gaze Imputation} \label{sec: wrist}
Given the known correlation between hand and eye movements~\cite{binsted2001eyehand, johansson2001eyehand} and the availability of wrist or hand motion data from common wearable devices, we extend \methodName{} by incorporating wrist movements as an additional modality. This experiment examines whether wrist information can further improve gaze imputation beyond head‑assisted modelling.

\subsection{Datasets}
The Nymeria dataset~\cite{ma2024nymeria} provides synchronised wrist motion data from two wristbands, making it suitable for our experiments. We use a subset containing 145.1h of training data, 8.3h for validation, and 28.7h for testing, after excluding recordings without valid wrist signals. Given its scale and the strong generalisation of \methodName{} observed in earlier experiments, all evaluations in this section are conducted on Nymeria.

\subsection{Evaluation Settings}
We adopted the same input duration, data loss ratios, and evaluation metrics as described in Section~\ref{sec: exp}. 
Since the dataset used here remains largely unchanged, we compared our method only against \methodNameOld~\cite{jiao25_uist}, which achieved the second-best performance in the previous experiments.  

To evaluate performance across different real-world usage scenarios, we trained several variants of \methodName{}: head only, head + one wrist, head + two wrists.
All models were trained using the same hyperparameters as in Section \ref{sec: exp}.

\subsection{Gaze Imputation Results}
Since all compared methods are diffusion‑based and can model gaze velocity distributions, we report only the MAE, as shown in Table \ref{tab: mae_nymeria_wrist}.  

Incorporating wrist information further improves performance, especially under high missing ratios. For example, at 90\% data loss, using both wrists yields a 3.5\% improvement over the head‑only variant. When the missing ratio is low (e.g., 10\%), all variants perform similarly. This suggests that eye–head coordination dominates short gaps (e.g., within typical blink durations), whereas eye–hand coordination becomes more informative over longer missing intervals.

\begin{table}[]
\centering
\begin{tabular}{lcrrr}
\toprule
Nymeria        & \multicolumn{4}{c}{Data loss ratio}                           \\ \midrule
               & 10\%           & 30\%          & 50\%          & 90\%          \\ \midrule
\methodNameOld            & 4.01 & 4.94 & 6.34 & 9.17 \\ 
\methodName              & {\ul 3.53} & {\ul 4.41} & 5.67 & 8.17 \\ 
\methodName + Left Wrist            & \textbf{3.52} & \textbf{4.34} & {\ul 5.52} & {\ul 7.98} \\ 
\methodName + Right Wrist            & 3.61 & 4.48 & 5.68 & 8.09 \\ 
\methodName + Two Wrists           & 3.54 & \textbf{4.34} & \textbf{5.49} & \textbf{7.88} \\ \bottomrule
\end{tabular}
\caption{MAE across different methods and data loss ratios on the Nymeria (Wrist) test set \cite{ma2024nymeria}. The best results are marked in bold, and the second-best are underlined.} \label{tab: mae_nymeria_wrist}
\end{table}

\section{Experiments – Gaze Generation}

When the data loss ratio reaches 100\%, the gaze imputation task becomes \textit{gaze generation}. As described in Section~\ref{sec: method_generation}, \methodName{} can be adapted to this setting with only minor changes. \change{Generating gaze directly from head motion and optional wrist/hand motion captured by commodity wearables also offers a practical solution for XR devices such as the Meta Quest~3, which lack built‑in eye tracking.}
Under the same 100\% missing-data setting, we also evaluate whether the generated 3D gaze directions can support egocentric gaze estimation when projected onto the image plane.

\subsection{Three-point Gaze Generation}\label{sec: generation_results}

\paragraph{Evaluation Settings}
We evaluate \methodName{} for gaze generation using the dataset introduced in Section~\ref{sec: wrist}.
We used the same input duration and evaluation metrics as in Sections~\ref{sec: exp} and~\ref{sec: wrist}, with the data loss ratio set to 100\%. To examine different practical configurations, we trained three variants of \methodName{}: head only, head + one wrist, and head + two wrists. All models were trained using the same hyperparameters as in Section~\ref{sec: exp}.

We compared \methodName{} with:

\begin{itemize}[leftmargin=*]
    \item \textbf{Pose2Gaze}~\cite{hu24pose2gaze}: Predicts gaze from head orientation and full-body poses using a CNN and GCN, and has been shown to outperform prior head‑only gaze‑generation methods~\cite{hu2020dgaze, hu2021fixationnet}. Note that full-body pose capture remains challenging in real-world settings.
\end{itemize}

We did not include HOIGaze~\cite{hu25hoigaze}, as it requires precise hand gestures and the 3D data of objects, which are not available in Nymeria and are difficult to obtain with commodity wearables.

\begin{table}[]
\centering

\begin{tabular}{lrr}
\toprule
Nymeria        & MAE & JS                           \\ \midrule

Pose2Gaze \cite{hu24pose2gaze}            & 13.09 & 0.238 \\ \midrule
\methodName              & {11.65} & {\ul 0.138}  \\ 
\methodName + Left Wrist            & 11.28 & 0.153  \\ 
\methodName + Right Wrist            & {\ul 10.98} & 0.156 \\ 
\methodName + Two Wrists           & \textbf{10.79} & \textbf{0.064} \\ \bottomrule
\end{tabular}
\caption{MAE and JS of gaze generation across different methods on the Nymeria (Wrist) test set \cite{ma2024nymeria}. The best results are marked in bold, and the second-best are underlined.} \label{tab: generation_nymeria_wrist}
\end{table}

\paragraph{Quantitative results.}
The quantitative results are presented in Table~\ref{tab: generation_nymeria_wrist}. 
All variants of \methodName{} achieved lower MAE and JS divergence than Pose2Gaze, indicating more accurate and more human-like gaze generation. 
The three-point variant (Head + Two Wrists) performs best, improving MAE by 17.6\% over Pose2Gaze (10.79 vs.\ 13.09) and achieving the lowest JS divergence (0.064). This shows that jointly leveraging eye--head--hand coordination enables \methodName{} to generate more accurate and biologically plausible gaze trajectories.

The two-point variants (Head + Left/Right Wrist) yield slightly lower MAE than the head-only model but higher JS divergence, suggesting that wrist motion provides useful spatial cues while introducing minor velocity distributional noise. Overall, these results confirm that multi-point motion signals from the head and wrists offer complementary information for generating realistic and temporally consistent gaze trajectories.

\begin{figure*}[!h]
    \centering
    \includegraphics[width=1.5\columnwidth]{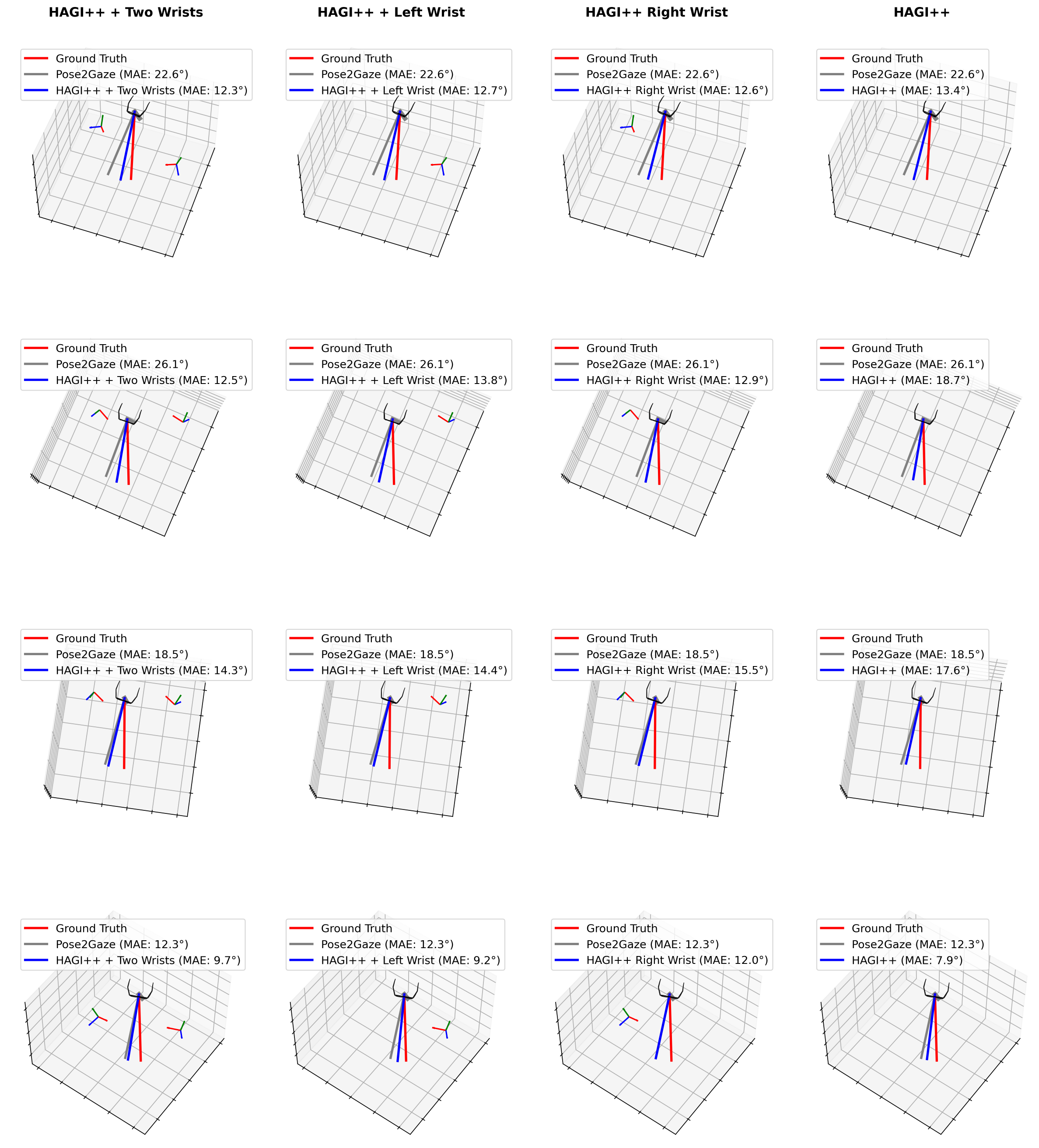}
    \caption{Visualisation of gaze generation results from the one-point, two-point, and three-point variants of \methodName{}, compared with the state-of-the-art Pose2Gaze~\cite{hu24pose2gaze}, which requires full-body pose input. Our method achieves higher accuracy using only signals from commodity wearable devices. Additional video results are provided in the supplementary materials.} 
    \label{fig:qual_gen}
\end{figure*}
\paragraph{Qualitative results.}
Figure~\ref{fig:qual_gen} visualises the gaze generation results of different \methodName{} variants alongside Pose2Gaze, which requires full-body pose input. All variants of our method achieve higher accuracy using only signals from commodity wearable devices. In the first three rows, the three-point \methodName{} yields the lowest MAE, indicating strong eye--hand--head coordination during these periods. In contrast, in the last row, the head-only variant performs best, suggesting that head movements dominate the coordination pattern at that time. Additional randomly selected qualitative video examples are provided in the supplementary materials.

\paragraph{Wrist Rotation vs. Translation.}
The results above show that adding wrist motion improves \methodName{} beyond the head-only model. To isolate the contribution of each component, we trained two ablations using only wrist rotations or only wrist translations.

As shown in Table~\ref{tab:wrist_ablation}, both variants reduce MAE relative to the head-only baseline, indicating that rotation and translation each provide useful cues. Their JS divergences remain similar, whereas combining both signals yields further accuracy gains and lowers JS divergence to 0.064. This demonstrates that wrist rotation and translation offer complementary information that enhances eye--head--hand coordination modelling in \methodName{}.

\begin{table}[]
\centering
\begin{tabular}{lrrrr}
\toprule
                  & \multicolumn{2}{c}{Wrist movements} & MAE                                  & JS                                                                    \\
                  & Rotation        & Translation       &          &              \\ \midrule
{\methodName} & \XSolidBrush               & \XSolidBrush               & 11.65          & 0.138                 \\
{\methodName + Two Wrists} & \XSolidBrush               & \Checkmark                 & 11.46          & 0.145                 \\
{\methodName + Two Wrists} & \Checkmark              & \XSolidBrush                 & {\ul 11.35}          & {\ul 0.129}                \\
{\methodName + Two Wrists} & \Checkmark               & \Checkmark                 & \textbf{10.79}          & \textbf{0.064}                 \\
 \bottomrule
\end{tabular}
\caption{The ablation study results of wrist rotation and translation on the Nymeria \cite{ma2024nymeria} (Wrist) test set. The best results are marked in bold, and the second-best are underlined.}\label{tab:wrist_ablation}

\end{table}

\change{\subsection{Egocentric Gaze Estimation}}

\change{Given the transformation between the gaze and camera coordinate systems, together with the camera intrinsics, the predicted 3D gaze directions can be projected onto the image plane for egocentric gaze estimation.}

\change{\paragraph{Evaluation Settings.}
Most existing egocentric gaze estimation methods~\cite{huang2018predicting, lai2022eye, lai2023eye} formulate the task as frame-wise gaze heatmap prediction and typically operate at relatively low frame rates (up to 8 Hz) due to their computational requirements. We therefore compared against EgoM2P~\cite{li2025egom2p}, a recent state-of-the-art foundation model that predicts 30 Hz gaze trajectories from 8 Hz egocentric video and has demonstrated superior performance over prior approaches.}

\change{Following the evaluation protocol of EgoM2P~\cite{li2025egom2p}, we use two-second input sequences and conduct a cross-dataset evaluation on Ego-Exo4D. \methodName{} was trained on the Nymeria dataset described in Section~\ref{sec: wrist}, with all training sequences reformatted to two-second windows, and evaluated on Ego-Exo4D without fine-tuning. Since Ego-Exo4d does not contain wrist motions, we only evaluated the head-only \methodName{}.}

\change{We report the mean angular error (MAE) of the predicted 3D gaze vectors. Following EgoM2P~\cite{li2025egom2p}, we additionally project both predicted and ground-truth gaze directions onto the image plane, normalise the coordinates to $[0,1]$, and report the mean squared error (MSE).}

\change{\paragraph{Results.}
\methodName{} outperforms EgoM2P on both evaluation metrics, achieving a lower MAE (\textbf{12.42}$^\circ$ vs.\ 13.47$^\circ$) and a lower MSE (\textbf{0.0075} vs.\ 0.0097). Notably, \methodName{} uses only head-motion signals captured by commodity wearable devices, whereas EgoM2P relies on egocentric video input. Despite not using any visual information, our method achieves superior performance, suggesting that head movements contain rich information for inferring gaze behaviour through eye--head coordination. These findings indicate that exploiting eye--head coordination can substantially reduce reliance on computationally intensive video modalities while maintaining accurate gaze estimation, making it a promising direction for efficient gaze-enabled XR systems.}

\section{Discussion}

\subsection{Performance} 
\paragraph{Gaze Imputation.}
Across all evaluations, \methodName{} consistently achieves lower MAE and produces gaze trajectories that more closely resemble natural human eye movements. It outperforms all baselines both quantitatively (Tables~\ref{tab: mae_nymeria}, \ref{tab: cross_dataset}) and qualitatively (Figure~\ref{fig:qual}). On average, \methodName{} reduces MAE by 24.45\% compared with the previous state of the art. Notably, its MAE at 30\% missing data is lower than baseline MAEs at 10\%, and its MAE at 50\% missing data is comparable to baselines at 30\%, indicating that \methodName{} can tolerate roughly 20\% more missing data while maintaining similar accuracy. Moreover, as shown in Table~\ref{tab: memory}, \methodName{} is more memory-efficient and faster at inference, making it well-suited for real-world mobile eye-tracking applications.

\paragraph{Gaze Generation.}
\methodName{} can also be applied to gaze generation. As shown in Section~\ref{sec: generation_results}, all variants of \methodName{} outperform prior state-of-the-art methods that rely on full-body motion capture, achieving MAE improvements from 11.0\% (head only) to 17.6\% (three-point) and lower JS divergence. This demonstrates that \methodName{} can synthesise accurate and biologically plausible gaze trajectories using only head and wrist signals—inputs readily obtainable from commodity wearables such as XR headsets and smartwatches—making it a practical solution for everyday gaze generation scenarios.
When the generated 3D gaze directions are projected for egocentric gaze estimation, \methodName{} also outperforms EgoM2P using only head-motion input, suggesting that eye--head coordination provides strong cues even without egocentric video.

\subsection{Head Rotation vs. Translation}
To examine the separate contributions of head rotation and translation, we conducted an ablation study. As shown in Table~\ref{tab:head_ablation}, both components improve gaze imputation, but head rotation has a notably stronger effect. This aligns with findings in vestibular research: dynamic visual acuity is primarily governed by the rotational vestibulo-ocular reflex (VOR), while translational VOR is less effective \cite{ramaioli2019vestibulo}. These results confirm that head rotation plays a more dominant role in eye--head coordination.

\subsection{Eye--Hand--Head Coordination}
As shown in Table~\ref{tab: mae_nymeria_wrist}, adding wrist information yields little MAE improvement at low missing ratios (10\%–30\%), where performance remains similar to the head-only variant. This suggests that over short time windows, gaze is primarily coupled with head motion, while hand motion contributes minimally. As the missing duration increases, wrist signals become more informative, leading to clear gains at 90\% and 100\% missingness (Tables~\ref{tab: mae_nymeria_wrist} and~\ref{tab: generation_nymeria_wrist}). Under the 100\% setting, using both wrists improves both MAE and JS divergence over the head-only model. Our ablation study further shows that wrist rotation and translation each provide complementary cues. Together, these findings highlight the synergistic role of eye--head--hand coordination, which \methodName{} effectively exploits to generate more accurate and natural gaze behaviour.

\subsection{Generalisability and Application}
As shown in Section~\ref{sec:results}, \methodName{} maintains its MAE in cross-dataset evaluations, indicating strong generalisability and enabling deployment without retraining when SLAM poses are available. Even with only head rotation, \methodName{} still outperforms existing methods, making it suitable for mobile eye trackers that rely solely on IMU data. Although real-world data include natural 0--50\,ms head–gaze delays, \methodName{} remains robust, underscoring its practicality.

\methodName{} supports three key applications: (1) post-processing to impute missing gaze samples for downstream machine learning models; (2) synthesising gaze for devices without eye tracking, including digital human animation and egocentric gaze estimation; and (3) imputing missing values at arbitrary positions within a 5-second window, enabling continuous correction after an initial warm-up. While current inference speed limits real-time use (Table~\ref{tab: memory}), future optimisation will enable applications such as gaze-based selection, foveated rendering, and attention-aware interfaces.

\subsection{Limitations and Future Work}
\methodName{} is evaluated only on egocentric mobile eye-tracking datasets. Although this reflects common real-world use, future work will explore its applicability to desktop setups with stationary eye trackers. Our method operates at 30\,Hz, while higher-frequency imputation is feasible given high-rate head-tracking sensors, this would require retraining and is currently limited by available datasets. \revise{In addition, although improving over the prior method \cite{hu24pose2gaze}, gaze generation under the challenging 100\% missing setting still shows a considerable gap to the ground truth, and further improving its prediction accuracy remains important for future work.}

Although our task assumes synchronised head and eye data, the datasets used contain natural 0--50\,ms delays, and \methodName{} remains robust. Future work will examine tolerance to larger misalignments common in practical systems.

Finally, we rely solely on motion signals from wearable devices. Since modern mobile eye trackers also provide egocentric video, and gaze is strongly influenced by visual context, we plan to integrate video information to jointly model scene content and body motion.

\section{Conclusion}
We introduced \methodName{}, a multi-modal diffusion-based approach for gaze imputation that leverages eye–head coordination to reconstruct missing gaze samples in a biologically plausible manner. Across three large-scale egocentric datasets, Nymeria, Ego-Exo4D, and HOT3D, \methodName{} consistently outperforms interpolation and state-of-the-art deep learning baselines, achieving up to a 25.3\% MAE reduction and producing more natural gaze velocity distributions. By incorporating wrist motion from commodity wearables, \methodName{} also extends to gaze generation and surpasses full-body motion–based methods; the generated gaze directions further support egocentric gaze estimation, outperforming EgoM2P using only head-motion input. Overall, our results highlight the value of modelling eye–head–hand coordination and establish \methodName{} as a robust and practical solution for improving gaze data quality in real-world and XR settings.


\ifCLASSOPTIONcompsoc
  \section*{Acknowledgments}
\else
  \section*{Acknowledgment}
\fi

This project has received funding from the European Union's Horizon Europe research and innovation funding program under grant agreement No. 101072410 and the National Natural Science Foundation of China (No. 62602542).

\ifCLASSOPTIONcaptionsoff
  \newpage
\fi

{\bibliographystyle{IEEEtran}
\bibliography{reference}}

\section{Biography Section}
\begin{IEEEbiographynophoto}{Chuhan Jiao}
is a PhD student at the University of Stuttgart, Germany. He received his BEng. in Computer Science and Technology from Donghua University, China, in 2020 and MSc. in Computer Science from Aalto University, Finland, in 2022. His research interest lies at the intersection of computer vision and human-computer interaction. 
\end{IEEEbiographynophoto}
\vspace{-20pt}

\begin{IEEEbiographynophoto}{Zhiming Hu} 
is a tenure-track Assistant Professor at The Hong Kong University of Science and Technology (Guangzhou), with a joint appointment at The Hong Kong University of Science and Technology. He was a post-doctoral researcher in the University of Stuttgart, Germany from August 2022 to July 2025.
He obtained his Ph.D. degree in Computer Software and Theory from Peking University, China in 2022 and received his Bachelor's degree in Optical Engineering from Beijing Institute of Technology, China in 2017. His research interests include virtual reality, human-computer interaction, eye tracking, and human-centred artificial intelligence.
\end{IEEEbiographynophoto}
\vspace{-20pt}

\begin{IEEEbiographynophoto}{Andreas~Bulling}
is Full Professor of Computer Science at the University of Stuttgart, Germany, where he directs the research group "Collaborative Artificial Intelligence". He received his MSc. in Computer Science from the Karlsruhe Institute of Technology, Germany, in 2006 and his PhD in Information Technology and Electrical Engineering from ETH Zurich, Switzerland, in 2010. Before, Andreas Bulling was an Independent Research Group Leader and Senior Researcher at the Max Planck Institute for Informatics as well as Saarland University, Germany. His research interests include computer vision, machine learning, and human-computer interaction.
\end{IEEEbiographynophoto}

\appendices
\section{Background}

\subsection{Gaze Data Imputation} \label{sec:defination}
Let $\mathbf{X}= \left\{x_1, x_2, ..., x_L \right\} \in \mathbb{R}^{L \times K}$ be a sequence of gaze directions, where $L$ represents the length of the gaze sequence, determined by the sampling rate of the eye tracker and the duration of data collection.
In our case, we set $K=2$, as each gaze direction at any given time step is represented by $(\text{pitch}, \text{yaw})$.
Additionally, we define $\mathbf{M}=\left\{m_1, m_2, ..., m_L \right\}  \in\{0,1\}^{L \times 1}$ to denote an observation mask, where $m_l = 1$ indicates that the eye tracker produces a valid output for $x_l$, while $m_l = 0$ denotes that $x_l$ is invalid.
Gaze data imputation is the task of estimating the gaze directions for invalid values within $\mathbf{X}$ by leveraging the valid gaze observations in $\mathbf{X}$.

\subsection{Conditional Score-based Diffusion Model for Time-Series Imputation (CSDI)} \label{sec:csdi}

Denoising Diffusion Probabilistic Models (DDPMs)~\cite{ho2020denoising} are a class of generative models that have achieved state-of-the-art performance across a range of domains, including image generation~\cite{ho2020denoising}, audio synthesis~\cite{kong2021diffwave}, and more recently, eye movement synthesis~\cite{jiao2024diffeyesyn, jiao2024diffgaze}. These models learn to generate realistic data by reversing a gradual noising process, enabling fine-grained control over the generative trajectory.

Tashiro et al.~\cite{tashiro2021csdi} extended the diffusion framework to time-series imputation by proposing Conditional Score-based Diffusion for Imputation (CSDI). CSDI models the conditional distribution of missing values given the observed portions of the time series. It introduces a conditional training scheme in which a masking function stochastically selects observed and unobserved regions of the input during training, allowing the model to learn flexible imputation strategies across various missing patterns.

Since gaze data is also time-series, CSDI can be adapted for gaze data imputation.
CSDI is trained in a self-supervised manner.
During training, given an input time series $\mathbf{x}_0$, CSDI randomly generates an observation mask that separates $\mathbf{x}_0$ into the observed part $\mathbf{x}^{co}_0$ and the target part requiring imputation $\mathbf{x}^{ta}_0$.
As with the original DDPM, CSDI consists of two processes: The forward process is a Markov chain that progressively adds noise to $\mathbf{x}^{ta}_0$, to transform $\mathbf{x}^{ta}_0$ into random noise following a Gaussian distribution. The forward process is defined as follows:
\begin{equation}
    q\left(\mathbf{x}^{ta}_t \mid \mathbf{x}^{ta}_0\right)=\mathcal{N}\left(\mathbf{x}^{ta}_t ; \sqrt{\bar{\alpha}_t} \mathbf{x}^{ta}_0,\left(1-\bar{\alpha}_t\right) I\right)
\end{equation}
where $t$ denotes the time step, $\bar{\alpha}_t = \prod_{s=1}^t \alpha_s$ and ${\alpha}_s$ is a constant determined by a predefined noise schedule. More specifically, 
\begin{equation}\label{eq:add_noise}
    \mathbf{x}^{ta}_t = \sqrt{{\bar{\alpha}_t}} \mathbf{x}^{ta}_0 + (1-\bar{\alpha}_t)\epsilon
\end{equation}
where $\epsilon \sim \mathcal{N}(0, I)$ is random Gaussian noise.

The reverse process aims to start from pure Gaussian noise, similar to $\mathbf{x}^{ta}_t$, and iteratively denoise it to reconstruct a sample resembling the original data distribution.
Since, in imputation, we additionally have conditional information from the observed sequence $\mathbf{x}^{co}_0$, the reverse process is defined as follows:
\begin{equation} \label{eq:reverse}
    p_\theta\left(\mathbf{x}^{ta}_{t-1} \mid \mathbf{x}^{ta}_t, \mathbf{x}^{co}_0\right)=\mathcal{N}\left(\mathbf{x}^{ta}_{t-1}  ; \mu_\theta(\mathbf{x}^{ta}_t, t \mid \mathbf{x}^{co}_0),\sigma_\theta(\mathbf{x}^{ta}_t, t \mid \mathbf{x}^{co}_0) I\right)
\end{equation}
where $\theta$ represents the trainable parameters of the neural network, 
\begin{equation} \label{eq:parameterization}
    \mu_\theta(\mathbf{x}^{ta}_t, t \mid \mathbf{x}^{co}_0) = \frac{1}{{\sqrt\alpha_t}}\left(\mathbf{x}^{ta}_t-\frac{1-\alpha_t}{\sqrt{1-\bar{\alpha}_t}} {\epsilon}_\theta\left(\mathbf{x}^{ta}_t, t \mid \mathbf{x}^{co}_0 \right)\right),
\end{equation}
and $\sigma_\theta(\mathbf{x}^{ta}_t, t \mid \mathbf{x}^{co}_0) I$ is a constant defined by the noise schedule. The term ${\epsilon}_\theta\left(\mathbf{x}^{ta}_t, t \mid \mathbf{x}^{co}_0 \right)$ in Equation \ref{eq:parameterization} is a trainable denoising deep learning model. 
The training objective is to minimise the difference between the prediction and the actual added noise in Equation \ref{eq:add_noise}:
\begin{equation}
    \mathcal{L}_{noise} = {\left\|{\epsilon_t - {\epsilon}_\theta\left(\mathbf{x}^{ta}_t, t \mid \mathbf{x}^{co}_0 \right)}\right\|}_2
\end{equation}

\section{Method}
We provide summarisations of \methodName{} training and inference in Algorithm \ref{alg:training} and \ref{alg:inference}, respectively.
\begin{algorithm}[h]
\caption{\methodName Training Procedure}\label{alg:training}
\begin{algorithmic}[1]
    \State \textbf{Input}: gaze data $\mathbf{X}$, corresponding \change{head movements and any available wrist/hand movements} $\mathbf{B}$, total diffusion step $T$
    
    \Repeat
        \State $\mathbf{x}_0$ = $\text{sin}(\mathbf{X})$
        \State $\mathbf{M} \sim \text{Random Mask Generator}$
        \State $\mathbf{x}_0^{co}$ = $\mathbf{M} \odot \mathbf{x}_0$
        \State $\mathbf{x}_0^{ta}$ = $(1 - \mathbf{M}) \odot \mathbf{x}_0$
        \State \(t \sim \text{Uniform}(\{1,...,T\})\)
        \State \(\epsilon \sim \mathcal{N}(0, I)\)
        \State \( \mathbf{x}_t^{ta} =\sqrt{\bar{\alpha}_t} \mathbf{x}^{ta}_0 + (1-\bar{\alpha}_t)\epsilon\)
        \State $\mathcal{L} = {\left\|{\epsilon_t - {\epsilon}_\theta\left(\mathbf{x}^{ta}_t, t \mid \mathbf{x}^{co}_0, \mathbf{B} \right)}\right\|}_2 $
        \State \(\nabla_\theta\mathcal{L}\)
    \Until {converged}
\end{algorithmic}
\end{algorithm}

\begin{algorithm}[h]
\caption{\methodName Sampling Procedure}\label{alg:inference}
\begin{algorithmic}[1]
    \State \textbf{Input}: gaze data $\mathbf{X}$, corresponding \change{head movements and any available wrist/hand movements} $\mathbf{B}$, observation mask $\mathbf{M}$
    \State $\mathbf{x}_0^{co}$ = $\mathbf{M} \odot \text{sin}(\mathbf{X})$
    \State Sample $\mathbf{x}^{ta}_T \sim \mathcal{N}(0, I)$
    \For{$t = T, T-1, \dots, 1$}
         \State \(\mu_\theta(\mathbf{x}^{ta}_t, t \mid \mathbf{x}^{co}_0, \mathbf{B}) = \frac{1}{\sqrt{\alpha_t}}\left(\mathbf{x}^{ta}_t-\frac{1-\alpha_t}{\sqrt{1-\bar{\alpha}_t}} {\epsilon}_\theta\left(\mathbf{x}^{ta}_t, t \mid \mathbf{x}^{co}_0,\mathbf{B} \right)\right)\)
         \State Sample $\mathbf{x}^{ta}_{t-1} \sim p_\theta\left(\mathbf{x}^{ta}_{t-1} \mid \mathbf{x}^{ta}_t, \mathbf{x}^{co}_0, \mathbf{B}\right)$ using Equation 3 in the main paper
    \EndFor
    \State \textbf{Output}: $\text{arcsin}(\mathbf{x}^{ta}_0)$
\end{algorithmic}
\end{algorithm}

\section{Gaze Imputation Experiment}

\subsection{Implementation details of baselines}
We provide details of our baseline selections and implementation.

\begin{itemize}[leftmargin = *]
    \item \textbf{Head direction}: Given the strong correlation between head and eye movements, head direction was widely used as a proxy for gaze direction in prior work \cite{hu2020dgaze, hu24pose2gaze, nakashima2015saliency, sitzmann2018saliency}.
    
    \item \textbf{Linear interpolation} and \textbf{Nearest interpolation}: These two interpolation techniques are the most commonly used methods for handling missing values in eye-tracking research \cite{kasneci2024introduction, grootjen2024uncovering}.
    
    \item \textbf{iTransformer} \cite{liu2024itransformer}: A Transformer-based model designed for multiple time-series tasks, including time-series imputation.
    
    \item \textbf{DLinear} \cite{Zeng2022AreTE}: A multilayer perceptron (MLP)-based method that models time-series trends using a moving average kernel and a seasonal component.
    
    \item \textbf{TimesNet} \cite{wu2023timesnet}: Converts 1D time-series data into 2D tensors with Fast Fourier Transformation and processes them using a CNN-based architecture. It was designed for multi-time-series tasks.
    
    \item \textbf{BRITS} \cite{cao2018brits}: A bidirectional RNN-based approach for time-series imputation.
    \item \textbf{CSDI} \cite{tashiro2021csdi}: A diffusion-based generative model for time-series imputation, demonstrating superior performance over GAN-based \cite{miao2021generative} and VAE-based \cite{fortuin2020gp} methods in various imputation tasks.
\end{itemize}

Since CSDI \cite{tashiro2021csdi} is also a diffusion-based approach, we ensured a fair comparison by training CSDI with the same diffusion hyperparameters as \methodName using its official implementation.
For other deep learning-based time-series imputation baselines, we leveraged PyPOTS \cite{du2023pypots}, a popular Python toolbox that includes various time-series imputation methods \cite{du2024tsibench}.
We modified only the input shape while keeping the optimal hyperparameters provided for the PhysioNet2012 dataset \cite{silva2012predicting} and trained all models with the same number of epochs and batch size.
All deep learning-based methods were trained on the Nymeria training set, and we selected the best-performing model on the validation set for the final evaluation.
For classical interpolation methods, we used the built-in functions from the SciPy library \cite{2020SciPy-NMeth}. As a baseline head direction proxy, we filled in missing frames with $(\text{pitch}, \text{yaw}) = (0, 0)$.
For generative models that do not produce deterministic outputs, we followed \cite{tashiro2021csdi, alcaraz2022diffusionbased} and used the median of 100 generated samples for evaluation.

\subsection{Result of time-series imputation baselines on Ego-Exo4d and HOT3D} \label{appendix}

We provide MAE of iTransformer \cite{liu2024itransformer}, Dlinear \cite{Zeng2022AreTE}, TimesNet \cite{wu2023timesnet}, BRITS \cite{cao2018brits} in gaze imputation task on the Ego-Exo4D \cite{grauman2024ego} and HOT3D \cite{banerjee2024hot3d} datasets in Table \ref{appendix}. 

\begin{table}[h]
\resizebox{\columnwidth}{!}{
\begin{tabular}{lrrrrrrrr}
             & \multicolumn{4}{c|}{Ego-Exo4D }                                                                             & \multicolumn{4}{c}{HOT3D }                        \\
             & \multicolumn{1}{c}{10\%} & \multicolumn{1}{c}{30\%} & \multicolumn{1}{c}{50\%} & \multicolumn{1}{c|}{90\%} & \multicolumn{1}{c}{10\%} & 30\%  & 50\%  & 90\%  \\ \hline
iTransformer \cite{liu2024itransformer} & 7.89                     & 11.04                    & 17.55                    & 26.36                     & 6.81                     & 9.83  & 15.80 & 24.46 \\
DLinear \cite{Zeng2022AreTE}      & 12.12                    & 12.14                    & 12.88                    & 14.00                     & 9.83                     & 10.00 & 10.62 & 11.52 \\
TimesNet \cite{wu2023timesnet}     & 21.66                    & 20.27                    & 22.37                    & 25.23                     & 19.83                    & 18.79 & 20.51 & 23.23 \\
BRITS  \cite{cao2018brits}      & 10.51                    & 12.47                    & 14.96                    & 19.17                     & 9.47                     & 11.36 & 13.50 & 17.20 \\ \hline
\end{tabular}}
\caption{Mean angular error (MAE) of iTransformer \cite{liu2024itransformer}, Dlinear \cite{Zeng2022AreTE}, TimesNet \cite{wu2023timesnet}, BRITS \cite{cao2018brits} on the Ego-Exo4D \cite{grauman2024ego} and HOT3D \cite{banerjee2024hot3d} datasets.} \label{table:appendix}
\end{table}

\subsection{\change{Temporal analysis of head orientation and gaze prediction}}

\begin{figure}[!h]
    \centering
    \includegraphics[width=\columnwidth]{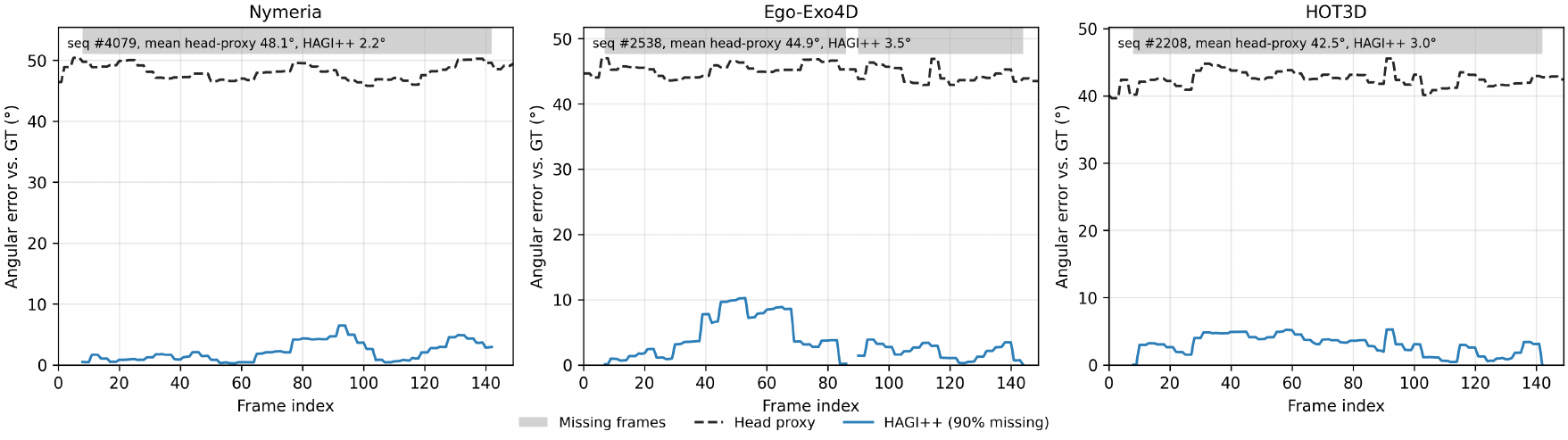}
    \caption{\change{\methodName Gaze prediction vs. head orientation over time on representative test sequences (90\% missing). x-axis: frame index. y-axis: angular error vs. GT (°). Dashed black: head orientation vs. GT. Blue: \methodName prediction vs. GT (missing frames). Gray bar: missing frames.} }
    \label{fig:head_gaze_temp}
\end{figure}

\change{We further analyse the relationship between gaze and head orientation over time. For each dataset, we select a representative test sequence with a large head--gaze discrepancy (mean head-proxy error $>$ 40$^\circ$) and evaluate \methodName under 90\% missing data (extreme cases).
Figure~\ref{fig:head_gaze_temp} plots the MAE between the ground-truth gaze and the head orientation, as well as the MAE between the ground-truth gaze and \methodName predictions. 
Across all sequences, head orientation remains a poor proxy for gaze, exhibiting errors of approximately 40--50$^\circ$ throughout the clips. In contrast, \methodName maintains low prediction errors despite 90\% of the gaze samples being missing. These results further demonstrate that HAGI++ does not simply follow head orientation, but instead recovers gaze dynamics beyond head pose.}

\section{Gaze Generation Experiment}
\subsection{\change{Adding additional visual input to \methodName}} 
\change{We conducted a preliminary study to investigate whether incorporating time-aligned egocentric video can further improve \methodName{} for gaze estimation. Following the same protocol as Section~IV-B of the main paper, we extracted visual features from each frame using DINOv3 with a ViT-B/16 encoder~\cite{simeoni2025dinov3}. The resulting video features were integrated into \methodName{} in the same manner as the optional wrist-motion features, using cross-attention to model interactions between gaze, head, and visual representations.}

\begin{figure*}[h]
    \centering
    \includegraphics[width=\textwidth]{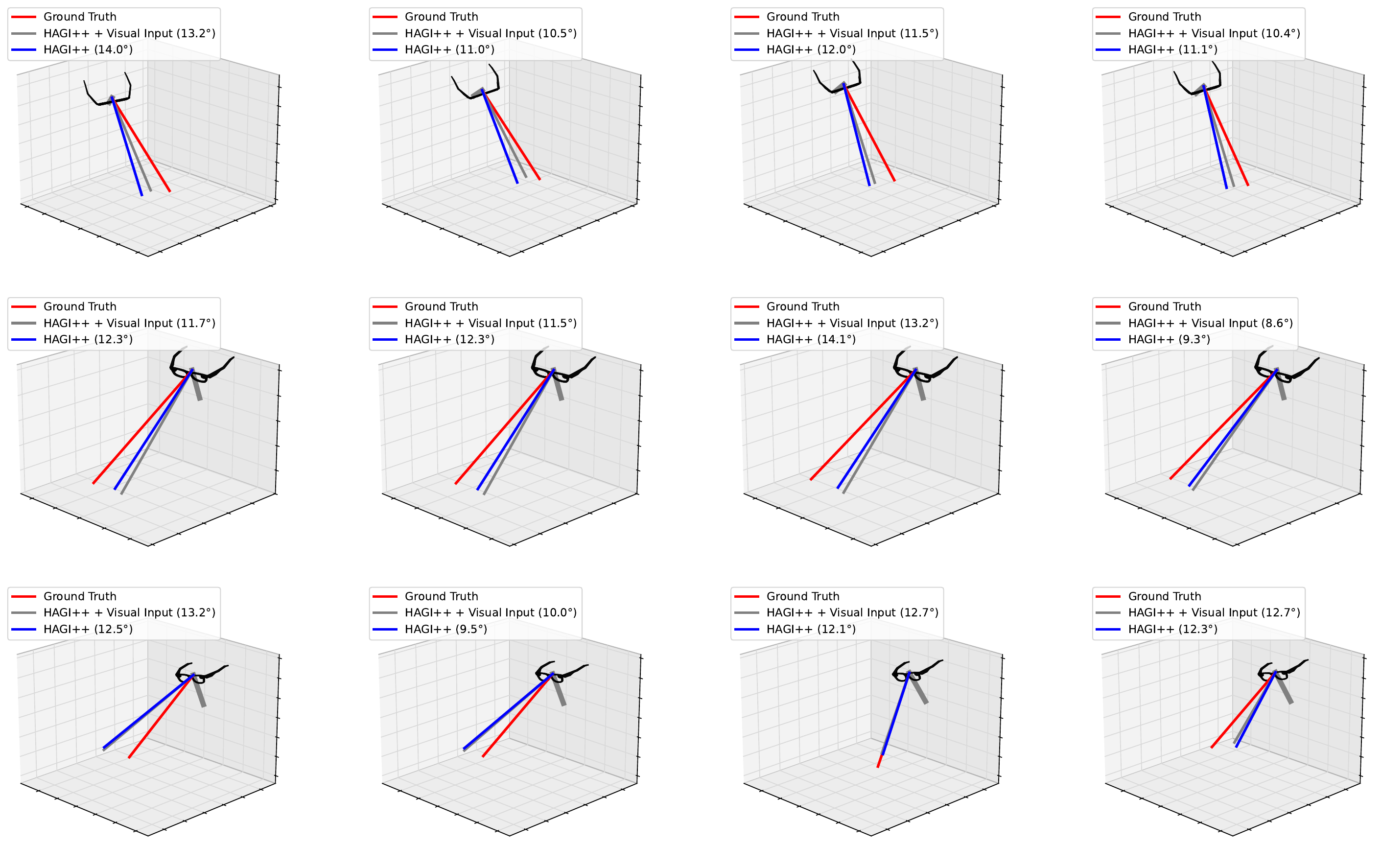}
    \caption{\revise{Visualisation of gaze prediction results from \methodName{} with and without visual input, compared against ground truth across four temporal frames in each sequence. The head-only variant uses eye--head coordination signals alone, while the visual-input variant additionally incorporates video features.}}
    \label{fig:qual_visual}
\end{figure*}

\change{As shown in Table~\ref{table:appendix_ego}, adding visual features improves both MAE (11.96$^\circ$ to 11.32$^\circ$) and MSE (0.0071 to 0.0068). However, the improvement is relatively modest, suggesting that eye--head coordination already provides substantial information for gaze prediction (\revise{See qualitative results in Figure \ref{fig:qual_visual}}). This finding is consistent with our main results, where head-motion signals alone achieve competitive performance across multiple tasks and datasets.}

\change{While visual information offers additional benefits, it also introduces considerable computational overhead due to video feature extraction and storage. These results indicate a trade-off between accuracy and efficiency: incorporating visual cues can further improve performance, whereas the head-only version of \methodName{} achieves comparable results with substantially lower computational requirements, making it more attractive for practical XR and wearable applications.} 

\begin{table}[h]
\centering
\begin{tabular}{lrr}

             MAE & MSE &  \\ \hline
\methodName  & 11.96                     & 0.0071  \\
\methodName + Visual Input  & \textbf{11.32}                     & \textbf{0.0068}  \\ \hline
\end{tabular}
\caption{Preliminary study on incorporating egocentric visual features into \methodName{} for gaze estimation on the Nymeria test set.} \label{table:appendix_ego}
\end{table}

\section{Discussion}
We provide extra discussion on the gaze imputation performance
\subsection{Gaze Imputation Performance}
Unlike MAE, which tends to increase with higher levels of missing data, the Jensen-Shannon (JS) divergence of \methodName{} \textit{decreases} as the missing ratio increases. We compute gaze velocity distributions using the \texttt{numpy.histogram} function with \texttt{bins=100} across all settings. However, the resulting absolute JS values are not directly comparable across missing ratios or datasets. For example, at 90\% missingness, the ground-truth velocity distribution is considerably broader than that at 10\%, with the latter effectively forming a subset of the former. Applying the same number of bins to both leads to differing bin widths, which in turn affects the scale of JS divergence. This makes direct cross-ratio or cross-dataset comparisons of JS scores inappropriate. Nonetheless, within each dataset and missing ratio, comparisons across methods remain valid. \methodName{} on par JS divergence with the state-of-the-art, indicating that it generates gaze trajectories with plausible velocity distributions.

As shown in Table I, existing time-series imputation models that are not based on diffusion perform worse than even standard interpolation techniques when applied to gaze data. This supports prior findings~\cite{jiao2024diffgaze} that diffusion models are better suited to modelling gaze velocity distributions. Non-diffusion methods often overfit to slow eye movements, such as fixations, which dominate real-world gaze recordings. While augmenting non-diffusion baselines with head movement information is technically feasible, it is unlikely to yield significant improvements, as the underlying distribution of gaze velocities remains unchanged. We therefore focus on diffusion-based approaches and demonstrate that integrating head input within this framework yields further improvements in performance.

\subsection{Limitation and Future Work}

\revise{Our current evaluation does not include a dedicated metric for temporal sequence similarity, as existing scanpath metrics are not directly applicable to our head-conditioned gaze generation setting. For example, Dynamic time warping can tolerate temporal misalignment, while Levenshtein distance requires discretising continuous gaze trajectories and ignores the magnitude of angular errors. Developing appropriate metrics that capture both temporal gaze dynamics and time-aligned eye--head coordination remains an important direction for future work.}

\vfill

\end{document}